\documentclass[lettersize,journal]{IEEEtran}
\usepackage{amsmath,amsfonts}
\usepackage{array}
\usepackage{textcomp}
\usepackage{stfloats}
\usepackage{url}
\usepackage[english]{babel}
\usepackage{ifthen}
\usepackage{multirow}
\usepackage{caption}
\usepackage{subcaption}
\usepackage{graphicx}
\graphicspath{{Figures/}}

\newcommand{\ing}[1]{\mathsf{#1}}
\newcommand{\Rn}[1]{\ifthenelse{\equal{#1}{}}{\mathbb{R}}{\mathbb{R}^{#1}}}
\newcommand{\Cn}[1]{\mathbb{C}^{\ing{#1}}}
\newcommand{\Rset}[2]{\ifthenelse{\equal{#2}{1}}{\in \Rn{#1}}{\in \Rn{#1 \times #2}}}
\newcommand{\Cset}[2]{\ifthenelse{\equal{#2}{1}}{\in \Cn{#1}}{\in \Cn{#1 \times #2}}}
\newcommand{\vect}[1]{\boldsymbol{\MakeLowercase{#1}}}
\newcommand{\mtrx}[1]{\boldsymbol{\MakeUppercase{#1}}}

\newcommand{\norm}[2]{\|#1\|_{#2}}

\newcommand{\transp}[1]{#1^{\mathsf{T}}}
\newcommand{\htransp}[1]{#1^{\mathsf{H}}}

\newcommand{\ind}[2]{\ifthenelse{\equal{#2}{}}{\chi_{#1}}{\chi_{#1}\left( #2 \right)}}

\newcommand{\iter}[2]{#1^{\ing{(#2)}}}
\DeclareMathOperator*{\argmin}{argmin}
\DeclareMathOperator*{\argmax}{argmax\,}

\usepackage{xcolor}
\newcommand{\revis}[1]{#1}

\usepackage{algorithm}
\usepackage{algpseudocode}
\algnewcommand\algorithmiccompute{\textbf{Compute:}}
\algnewcommand\Compute{\item[\algorithmiccompute]}
\algnewcommand\algorithmicassemble{\textbf{Assemble:}}
\algnewcommand\Assemble{\item[\algorithmicassemble]}
\algnewcommand\algorithmicmyreturn{\textbf{Return:}}
\algnewcommand\Myreturn{\item[\algorithmicmyreturn]}

\begin{document}

\title{Blind identification of Ambisonic reduced room impulse response}
%
\author{Sr\dj{}an Kiti\'c and J\'er\^ome Daniel \thanks{S.~Kiti\'c was with Orange Labs, France, at the time of writing this article. J.~Daniel is with Orange Labs, France. The two authors have equally contributed to the present article. \\
This paper has supplementary downloadable material available at http://ieeexplore.ieee.org., provided by the authors. The material includes experimental results complementing those presented in the article. Contact jerome.daniel@orange.com for further questions about this work.}}

%
\maketitle
\begin{abstract}
Recently proposed \emph{Generalized Time-domain Velocity Vector (GTVV)} is a generalization of relative room impulse response in spherical harmonic (\emph{aka} Ambisonic) domain, that allows for blind estimation of early-echo parameters: the directions and relative delays of individual reflections. However, the derived closed-form expression of GTVV mandates few assumptions to hold, most important being that the impulse response of the reference signal needs to be a minimum-phase filter. In practice, the reference is obtained by spatial filtering towards the Direction-of-Arrival of the source, and the aforementioned condition is bounded by the performance of the applied beamformer (and thus, by the Ambisonic array order). In the present work, we \revis{circumvent this problem by directly modeling the impulse responses constituting the GTVV time series}, which permits not only to relax the initial assumptions, but also to extract the information therein in a more consistent and efficient manner, entering the realm of blind system identification. Experiments using simulated and recorded room impulse responses confirm the effectiveness of the proposed approach.
\end{abstract}
\begin{IEEEkeywords}
blind identification, Ambisonic, microphone array, RTF, Prony
\end{IEEEkeywords}
\section{Introduction}
\label{sec:intro}

Room Impulse Response (RIR) can be thought of as an ``acoustic fingerprint'' of the surrounding environment \cite{kuttruff2016room}, and its importance in spatial audio processing cannot be overstated. It encodes the information about acoustic multipath - reverberation, which inevitably affects all indoor audio recordings. Traditionally often seen as a nuisance, reverberation is known to degrade the results of localization algorithms \cite{evers2020locata}, worsen automatic speech recognition (ASR) and intelligibility \cite{kinoshita2016summary}, and negatively impact sound source separation \cite{zhang2020end}. Nevertheless, a number of recent works has demonstrated that \emph{early} reverberation actually have a potential to improve performances in various tasks. These ``echo-enabled'' methods exploit early reflections to boost performance of acoustic localization \cite{o2010automatic,di2019mirage,daniel2020time}, source separation \cite{scheibler2018separake,weisman2021robustness,shmaryahu2022importance}, speech and sound event recognition \cite{giri2015improving,yasuda2022echo,ratnarajah2023towards}, but also to address some unconventional problems such as localization behind soundproof obstacles \cite{kitic2014hearing,an2018reflection,boger2023towards}, inference of room geometry \cite{dokmanic2013acoustic,lovedee2019three,antonacci2012inference}, distance estimation \cite{ribeiro2010using,birnie2019reflection,daniel2022echo}, identification of room acoustic parameters \cite{yu2020room,dilungana2022geometry,baum2022environment} and acoustic matching \cite{su2020acoustic,tang2020scene}.

While the availability of pre-recorded RIRs would be, therefore, very beneficial for many echo-enabled applications, such procedure demands specific equipment and skills. Moreover, a particular RIR is dependent on the given acoustic conditions, hence in dynamic scenes (\emph{e.g.}, when microphone or source are mobile) one would require repeated RIR measurements, which is clearly impractical. Thus, there is a growing interest for adaptive blind system identification (BSI) methods - subject to certain assumptions, these are capable of inferring RIRs (up to a common delay and scale \cite{chi2006blind}) using only recorded audio signals. However, classical adaptive BSI methods (\emph{e.g.}, multichannel least mean squares (MCLMS) algorithm \cite{xu1995least}), have notable drawbacks, such as their sensitivity to noise and incorrect model order \cite{naylor2010speech}.

Recently, we have investigated properties of the so-called Generalized Time-domain Velocity Vector (GTVV) \cite{kitic2022generalized}, a generalization of the well-known relative room impulse response in spherical harmonic (SH) domain \cite{jarrett2017theory}. The main advantage of GTVV over classical relative RIR is due to its reference signal, which is obtained by beamforming towards the (approximate) Direction-of-Arrival (DoA) of a far field source. Assuming that the reference signal is dominated by direct propagation, the GTVV representation admits a closed-form expression that can be used to directly infer the DoAs and relative delays of individual wavefronts (including the direct component). To satisfy this assumption, beamformer needs to be sufficiently selective, which is acceptable for Higher-Order Ambisonic (HOA) \cite{zotter2019ambisonics} arrays, but becomes prohibitive when prevalent \cite{lee2021multichannel} First Order Ambisonic (FOA) arrays are used. Indeed, the beam width of FOA beamformers is too permissive \cite{jarrett2017theory}, thus a number of non-attenuated reflections invalidates the former requirement. 


The main contribution of this work is a method for the identification of Ambisonic RIRs with scale and delay shifting according to the principal wave front, directly from the GTVV imprint (thus, blindly) and regardless of the array order. Hence, we term this representation Reduced Room Impulse Response (RdRIR), all the more that we are primarily interested in extracting the early part of Ambisonic RIRs containing directional information, i.e. the “early echoes”. 
Our goal is also to retain low computational complexity, as well as to facilitate implementation. Therefore, we compare several algorithmic variants having different levels of complexity, and evaluate their estimation performance. We show through experiments on simulated and 
real impulse response data that the proposed methods are effective in extracting parameters of multiple wavefronts, under various acoustic conditions. 


This work unifies and complements our previous contributions published as conference papers \cite{daniel2020time,kitic2022generalized}. The article is organized as follows: after a review of prior art given in Section~\ref{secRelated}, we proceed to the signal model behind GTVV in Section~\ref{secGTVV}, and discuss its estimation and limitations. This is succeeded by presenting the RdRIR estimation methods in Section~\ref{secRdRIR}.
The results of computer experiments are given in Section~\ref{secExperiments}. The article is concluded in Section~\ref{secConclusion}.

\textbf{Notations: }Real- or complex-valued scalar variables are written in lowercase italic or greek alphabet, while we use boldface font for vectors (lowercase) and matrices (uppercase). Serif font is reserved for integers, with the uppercase serif denoting constant integer values. For a matrix $\mtrx{M}$, we use $\vect{m}_{\ing{i},:}$ and $\vect{m}_{:,\ing{j}}$ to specify its $\ing{i}$\textsuperscript{th} row and $\ing{j}$\textsuperscript{th} column, respectively, and $m_{\ing{i,j}}$ to denote the matrix entry at their intersection. The sets of natural, real and complex numbers are denoted by $\mathbb{N}$, $\mathbb{R}$ and $\mathbb{C}$, respectively. The Fourier transform \revis{(or the Discrete Fourier Transform, where appropriate)} and its inverse are given by $\mathcal{F}$ and $\mathcal{F}^{-1}$, while a variable in frequency domain is marked by the circumflex accent. The transpose and conjugate transpose operations are denoted by $\transp{(\cdot)}$ and $\htransp{(\cdot)}$, respectively. The notation $\iter{x}{i}$ refers to the $\ing{i}$\textsuperscript{th} iterate of some algorithmic variable $x$. For a vector-valued function $\vect{x}(t)$, its corresponding \emph{delay-magnitude representation} is defined as $\zeta_{\vect{x}}(t) = \norm{\vect{x}(t)}{2}$.

\section{Prior art}
\label{secRelated}

Research in BSI has flourished since the seminal work of Sato \cite{sato1975method} on self-recovering equalization for digital communications. In the context of Single-Input-Multiple-Output (SIMO) systems, where the same source excites multiple channels (as in our problem setting), the concept of \emph{cyclostationarity} in second order statistics (SOS) \cite{tong1991new} was widely adopted. Different SOS variants have been proposed, based on channel cross-relation (CR) \cite{xu1995least}, subspace decomposition \cite{moulines1995subspace} and maximum likelihood estimation \cite{hua1996fast}. The CR technique has been particularly popular, and was later extended to adaptive BSI for acoustic channel identification, either in time \cite{benesty2000adaptive,huang2002adaptive}, or frequency domain \cite{huang2003class,ahmad2006proportionate}. Nevertheless, while such methods have evolved in order to improve their robustness to noise, in general they are known to perform well only under sufficiently high Signal-to-Noise Ratio (SNR) \cite{naylor2010speech}. \revis{Some CR variants have been tailored to estimation of early RIRs, the task referred to as ``under-modeled BSI'' in the literature \cite{xue2016cross,xue2017frequency}.} Contemporary approaches based on deep learning have only recently been employed for blind RIR estimation \cite{wager2020dereverberation,steinmetz2021filtered,ratnarajah2023towards}. Some of these models achieve impressive performance on different metrics and datasets, but are currently limited to predicting only single channel RIRs. The interpolation of missing RIR channels of a circular microphone array, in the SIMO context, has been formulated and solved as an inverse problem regularized by deep prior in \cite{pezzoli2022deep}. 

Relative Transfer Function (RTF) is a well-known concept in microphone array signal processing, that has been in widespread use for decades \cite{gannot2017consolidated}. There are two distinct conveniences of RTF: i) it is obtained using only received multichannel signals \revis{as their ratio in frequency domain}, and ii) it is theoretically a source signal-agnostic representation (thus, encoding only the propagating characteristics of the environment). Relative transfer functions have also been adopted in Ambisonic domain \cite{jarrett2017theory,perotin2018multichannel,bosca2021dilated}, yet sometimes under different names (\emph{e.g.} relative harmonic coefficients (RHC) \cite{hu2020semi} or frequency-domain velocity vector (FDVV) \cite{daniel2020time}). For FOA signals, its real part is aligned with \emph{pseudointensity} vector \cite{merimaa2006analysis,jarrett20103d}, a widely used alternative to steered beamforming for low-cost DoA estimation. The temporal representation of RTF is relative impulse response \cite{meier2015analysis,gannot2001signal} (again renamed to time-domain velocity vector (TDVV) in our earlier work \cite{daniel2020time}). An idea related to RTF and relative impulse response was discussed by G{\"o}lles and Zotter in \cite{golles2020directional}, where they calculate the ratio of received Ambisonic signals directly in time domain.

Classical RTF and relative impulse response have been extended to generalized (frequency and time domain) velocity vectors \cite{kitic2022generalized}, mentioned before and explained in detail in the next section\footnote{Note that, in \cite{herzog2021generalized}, Herzog and Habets have proposed another acoustic quantity termed \emph{generalized intensity vector}, which, despite the naming similarity, is different from generalized velocity vectors discussed here.}. The value of the beamformed reference has also been recognized in \cite{borrelli2018denoising,meyer2022blind}, where the authors take it to be the proxy for the source signal, hence the obtained ratio is considered an approximation of the acoustic transfer function (ATF). However, this hypothesis can only be valid if the applied beamformer filters out all reflections, which is rarely the case. To alleviate this issue, in \cite{meyer2022blind} the authors propose to use a time-frequency mask obtained by the improved direct-path-dominance test \cite{madmoni2018direction}, which is nonetheless a costly estimator in terms of computational complexity. In \cite{biderman2016efficient}, this representation has been used for denoising, under the assumption that the directions of acoustic reflections are known a priori.

As mentioned before, the central motivation of our work is extracting spatio-temporal information about the early echoes, and not identifying the complete propagation channels. A related work has been recently published by Shlomo and Rafaely \cite{shlomo2021blind}, where they propose the phase aligned spatial correlation (PHALCOR) algorithm, for the same purpose. They obtain convincing results on simulated data, at the expense of high computational cost and somewhat intricate implementation, involving singular vector decomposition, sparse analysis and clustering.

\section{Generalized Velocity Vector}
\label{secGTVV}

In this section we recall and extend the concept of generalized velocity vector, introduced in \cite{kitic2022generalized}. \revis{In particular, generalized velocity vector definition is broadened to include frequency-dependent beamforming and attenuation, and its relation with RTF and pseudointensity is made more explicit. Moreover, we provide a closed-form expression for GTVV for the case where the dominant wavefront in the reference signal is an acoustic reflection (instead of direct sound).}

\subsection{Signal model}

Let $\hat{\vect{b}}(f) \in \Cn{(L+1)^2}$ denote the vector of concatenated spherical harmonic expansion coefficients (the ``HOA channels'') up to order $\ing{L}$, at frequency $f$. We assume that mode strength compensation \cite{jarrett2017theory} has been applied, and that the recorded signals are due to a far field point source at azimuth $\theta_0$, elevation $\phi_0$ and range $d_0$ from the microphone array, in an indoor environment. We approximate $\hat{\vect{b}}(f)$ as follows:
\begin{align}
  \hat{\vect{b}}(f) \approx & \hat{s}(f) \hat{\chi}(f) \sum\limits_{\ing{n}=0}^{\ing{N}-1} \hat{\nu}_{\ing{n}}(f) \vect{y}_{\ing{n}} e^{-j2\pi f \bar{\tau}_n} + \hat{\vect{e}}(f) \\
  = & \hat{\vect{x}}(f) + \hat{\vect{e}}(f).
\end{align}

\revis{In the expression above, $\hat{s}(f)$ represents the source (excitation) wideband signal, such as speech, while $\hat{\chi}(f)$ is an anti-aliasing filter applied before the analog-to-digital converter. The terms} ${\hat{\nu}_{\ing{n}}(f) \in (0,1)}$, $\bar{\tau}_n \in \mathbb{R}^+$ and $\vect{y}_{\ing{n}} \in \Rn{(\ing{L}+1)^2}$ are the attenuation factor, Time-of-Arrival (ToA), and the real-valued SH encoding vector of the $\ing{n}$\textsuperscript{th} acoustic wavefront, respectively. The plane wave expansion has been truncated to $\ing{N}$ wavefronts, aggregating direct propagation and early echoes in $\hat{\vect{x}}(f)$, while the late reverberation and diffuse noise are represented by an additive term $\hat{\vect{e}}(f)$. The ToA of the direct signal is $\bar{\tau}_0 \approx d_0/c$ (where $c$ is the speed of sound), while $\bar{\tau}_{\max} = \max_{\ing{n}} \bar{\tau}_n$ (roughly) corresponds to the mixing time \cite{hidaka2007new} of the room. \revis{Note that there is an implicit dependency between $\ing{N}$, $\bar{\tau}_{\max}$ and the geometric and acoustic properties of the environment.} 

The diffuse component $\hat{\vect{e}}(f)$ is considered uncorrelated with the directional term $\hat{\vect{x}}(f)$ \cite{jarrett2017theory}. The latter is modeled by the image source model (ISM) \cite{allen1979image}, which considers all reflections to be specular, and approximates the frequency-dependent factor $\hat{\nu}(f)$ by absorption coefficient - an attenuation of the signal magnitude by a positive factor lower than $1$. 
In the ISM model, hence, the phase shifts of individual wavefronts are only due to differences in lengths of their acoustic paths. This is a limitation of the model - in general, $\hat{\nu}(f)$ is a complex variable that encodes the material absorption and air attenuation, and depends on the angle of incidence \cite{kuttruff2016room}.

Given a beamformer $\hat{\vect{w}}(f) \in \mathbb{C}^{(\ing{L+1})^2}$, constrained by $\htransp{\hat{\vect{w}}(f)}\vect{y}_0 = \beta_0 \in \mathbb{R}^+ = \text{const}$ (without loss of generality, we consider $\beta_0=1$), \emph{Generalized Frequency-domain Velocity Vector} (GFVV) has been defined \cite{kitic2022generalized} as 
\begin{multline}\label{eqFDVVinst}
  \hat{\vect{v}}(f) = \frac{\hat{\vect{x}}(f)}{\htransp{\hat{\vect{w}}(f)}\hat{\vect{x}}(f) } =  \frac{\vect{y}_0 + \sum\limits_{\ing{n}=1}^{\ing{N}-1} \hat{g}_{\ing{n}}(f) e^{-j2\pi f \tau_{\ing{n}} } \vect{y}_{\ing{n}} }{1 + \sum\limits_{\ing{n}=1}^{\ing{N}-1} \hat{g}_{\ing{n}}(f) \hat{\beta}_{n}(f) e^{-j2\pi f \tau_{\ing{n}} }}.
\end{multline}
Here, $\hat{g}_{\ing{n}}(f) = \hat{\nu}_{\ing{n}}(f)/\hat{\nu}_0(f) < 1$, $\tau_{\ing{n}} = \bar{\tau}_{\ing{n}} - \bar{\tau}_0 > 0$ and $\hat{\beta}_{\ing{n}}(f) = \htransp{\hat{\vect{w}}(f)}\vect{y}_{\ing{n}} / \beta_0$ denote the attenuation, delay and spatial response of the ${\ing{n}}$\textsuperscript{th} reflected plane wave \emph{relative} to the direct propagation component, respectively. 

We further assume that $\hat{\kappa}_{\ing{n}}(f) = \hat{g}_{\ing{n}}(f)\hat{\beta}_{\ing{n}}(f)$ is sufficiently smooth, such that its time domain counterpart $\kappa_{\ing{n}}(t)$ is compact\footnote{By the virtue of Paley-Wiener theorem \cite{rudin1991functional}.}. If $\hat{\vect{w}}(f):=\vect{w}$ is a wideband beamformer, this condition is usually satisfied, since $\hat{g}_{\ing{n}}(f)$ (which can be thought of as a scaled attenuation factor), is often a slowly-varying function of frequency in standard rooms \cite{kuttruff2016room}. However, care should be taken with some data-dependent beamformers, such as Minimum Power Distortionless Response (MPDR), which may exhibit abrupt changes in directivity \cite{rafaely2015fundamentals}.

Note that GFVV is (ideally) agnostic with regards to $\hat{s}(f)$ and $\hat{\chi}(f)$, hence, we can rewrite \eqref{eqFDVVinst} as:
\begin{equation}\label{eqGFVVcompact}
  \hat{\vect{v}}(f) = \frac{\hat{\vect{h}}(f)}{\hat{a}(f)},
\end{equation} 
where $\hat{\vect{h}}(f)$ is the numerator of the rightmost part of \eqref{eqFDVVinst}, while $\hat{a}(f)=\htransp{\hat{\vect{w}}(f)}\hat{\vect{h}}(f)$. The channel-wise inverse Fourier transform of GFVV, yields its temporal analogue, \emph{i.e.}, GTVV:
\begin{equation}\label{eqGTVVinst}
  \vect{v}(t) = \mathcal{F}^{-1} \left(\hat{\vect{v}}(f) \right) = \vect{h}(t) \ast a^{-1}(t),
\end{equation}
with $(a \ast a^{-1})(t) = \delta(t)$.

It is noteworthy that the standard RTF in SH domain \cite{jarrett2017theory} \revis{(\emph{i.e.}, FDVV)}, for which the reference is the first (omnidirectional) Ambisonic channel, is a special case of GFVV obtained by setting $\vect{w} = \transp{\left[ \begin{smallmatrix} 1 & 0 & 0 & \hdots & 0 \end{smallmatrix} \right]}$. Nonetheless, it would be preferential to use a beamformer steered approximately towards DoA of the source, as discussed later in this section. \revis{This also clarifies the notion of ``generality'' in GFVV - its reference channel does not correspond to one particular HOA channel, but is a linear combination of all available channels. Likewise, TDVV (relative impulse response in SH domain) becomes a special case of GTVV.} We remind the reader that the pseudointensity vector \cite{jarrett20103d, merimaa2006analysis} corresponds to the real part of RTF for the FOA signals. In \cite{daniel2020time}, we have argued that this classical DoA estimator is biased in the presence of strong reflections, but, without providing a detailed explanation. We take the opportunity to elaborate this claim in Appendix~\ref{appPIV}.




\subsection{GTVV estimation}\label{ssec:estimation}

The expression \eqref{eqFDVVinst} cannot be used directly, even in the noiseless setting, since we expect $\hat{\vect{e}}(f)$ to contain the diffuse late reverberation components. Moreover, a practical estimation method needs to be robust to low SNR levels. The following computationally efficient estimator has been proposed in \cite{kitic2022generalized,daniel2022echo}, and represents an adaptation of the well-known estimator based on speech signal nonstationarity \cite{shalvi1996system,gannot2001signal}. From \eqref{eqFDVVinst}, we have that a GFVV entry $\hat{v}_{\ing{l}} (f)$ could be seen as the ratio between ``denoised'' versions of $\hat{b}_{\ing{l}}(f)$ and the reference:
\begin{equation}
  \hat{v}_{\ing{l}} (f) = \frac{\hat{b}_{\ing{l}}(f) - \hat{e}_{\ing{l}}(f)}{\sum_{\ing{l}'} \hat{w}^*_{\ing{l}'}(f) \left(\hat{b}_{\ing{l}'}(f) - \hat{e}_{\ing{l}'}(f)\right)}.
\end{equation}
Rearranging the terms in the expression above gives
\begin{equation}
  \hat{b}_{\ing{l}}(f) = \hat{v}_{\ing{l}} (f) \sum_{\ing{l}'} \hat{w}^*_{\ing{l}'}(f) \hat{b}_{\ing{l}'}(f) + \hat{n}_{\ing{l}}(f)
\end{equation}
where we denote $\hat{n}_{\ing{l}}(f) = \hat{v}_{\ing{l}}(f) \sum_{\ing{l}'} \hat{w}^*_{\ing{l}'}(f)\hat{e}_{\ing{l}'}(f) - \hat{e}_{\ing{l}}(f)$. 

Since the two terms on the right hand side are correlated, we will estimate $\hat{v}_{\ing{l}}(f)$ and noise statistics simultaneously, in the least-squares sense, as originally proposed in \cite{shalvi1996system}. First, the signal is analyzed in time-frequency -- particularly, Short-time Fourier transform (STFT) -- domain:
\begin{equation}
  \hat{b}_{\ing{l}}(\ing{f,t}) = \hat{v}_{\ing{l}} (\ing{f,t}) \sum_{\ing{l}'} \hat{w}^*_{\ing{l}'}(\ing{f}) \hat{b}_{\ing{l}'}(\ing{f,t}) + \hat{n}_{\ing{l}}(\ing{f,t}),
\end{equation}
where $(\ing{f,t})$ designates discrete frequency and time frame indices, respectively, with a slight abuse of notation. Multiplying both sides by $b^*_{\ing{l}}(\ing{f,t})$, and taking expectation yields
\begin{equation}\label{eqCompact}
  \hat{\phi}_{\ing{l}}^2(\ing{f,t}) = \hat{v}_{\ing{l}} (\ing{f,t}) \sum_{\ing{l}'} \hat{w}^*_{\ing{l}'}(\ing{f})  \hat{\phi}_{\ing{l}',\ing{l}}(\ing{f,t}) + \hat{\sigma}_{\ing{l}}(\ing{f,t}),
\end{equation}
where $\hat{\phi}_{\ing{l}}^2(\ing{f,t}) = \mathbb{E}[|\hat{b}_{\ing{l}}(\ing{f,t})|^2]$ is the variance of the $\ing{l}$\textsuperscript{th} channel, while $\hat{\phi}_{\ing{l}',\ing{l}}(\ing{f,t}) = \mathbb{E}[\hat{b}_{\ing{l}'}(\ing{f,t})\hat{b}^*_{\ing{l}}(\ing{f,t})]$ and $\hat{\sigma}_{\ing{l}}(\ing{f,t})$ are the cross-correlations between channels $\ing{l}'$ and $\ing{l}$, and between the noise term $\hat{n}_{\ing{l}}$ and $\hat{b}_{\ing{l}}$, respectively. Under the assumption that the noise statistics $\hat{\sigma}_{\ing{l}}$ and the acoustics of the environment (thus, GFVV) evolve slower than speech statistics, and by rearranging the terms, the expression \eqref{eqCompact} is approximated by
\begin{align}\label{eqCompactApprox}
  \hat{\phi}_{\ing{l}}^2(\ing{f,t}) & \approx \left[ \begin{matrix} \htransp{\hat{\vect{\phi}}_{:,\ing{l}}(\ing{f,t})} \hat{\vect{w}}(\ing{f})   & 1  \end{matrix} \right] \left[ \begin{matrix} \hat{v}_{\ing{l}} (\ing{f}) \\ \hat{\sigma}_{\ing{l}}(\ing{f}) \end{matrix} \right], \; \text{where}\\
  \hat{\vect{\phi}}_{:,\ing{l}}(\ing{f,t}) & = \transp{\left[ \begin{smallmatrix} \hat{\phi}_{0,\ing{l}}(\ing{f,t}) && \hat{\phi}_{1,\ing{l}}(\ing{f,t}) && \hdots && \hat{\phi}_{(\ing{L+1})^2-1,\ing{l}}(\ing{f,t}) ] \end{smallmatrix}  \right]}.
\end{align}
The approximation above is assumed to hold for a collection of frames centered at $\ing{t}_0$, \emph{i.e.} within ${\ing{t} \in [\ing{t}_0 - \ing{T}/2, \ing{t}_0 + \ing{T}/2]}$, for $\ing{T} \in 2\mathbb{N}$. Note that $\hat{v}_{\ing{l}} (\ing{f})$ and $\hat{\sigma}_{\ing{l}}(\ing{f})$ are assumed constant for this set of frames, which is compactly written as 
\begin{align}\label{eqCompactApproxVect}
  \hat{\vect{\phi}}_{\ing{l}}^2(\ing{f}) & \approx \left[ \begin{matrix} \htransp{\hat{\mtrx{\Phi}}_{:,\ing{l}}(\ing{f})} \hat{\vect{w}}(\ing{f})   & \vect{1}  \end{matrix} \right] \left[ \begin{matrix} \hat{v}_{\ing{l}} (\ing{f}) \\ \hat{\sigma}_{\ing{l}}(\ing{f}) \end{matrix} \right], \; \text{where}\\
  \hat{\vect{\phi}}_{\ing{l}}^2(\ing{f}) &= \transp{\left[ \begin{smallmatrix} \hat{\phi}^2_{\ing{l}}(\ing{f,t_0 + T/2}) && \hdots && \hat{\phi}^2_{\ing{l}}(\ing{f,t_0 - T/2}) ] \end{smallmatrix}  \right]}, \label{eqPhiVect} \\
  \hat{\mtrx{\Phi}}_{:,\ing{l}}(\ing{f}) &= \left[ \begin{smallmatrix} \hat{\vect{\phi}}_{:,\ing{l}}(\ing{f,t_0 + T/2}) && \hdots &&  \hat{\vect{\phi}}_{:,\ing{l}}(\ing{f,t_0 - T/2}) \end{smallmatrix}  \right] \label{eqPhiMtx}
\end{align}
and $\vect{1}$ is the all-one vector. This is an overdetermined linear system that can be solved efficiently in the least squares sense (amounts to the inversion of a $2 \times 2$ matrix), providing an estimate of $\hat{v}_{\ing{l}} (\ing{f})$ at frame $\ing{t_0}$.

The quality of the GFVV estimate depends on a number of factors, including the STFT parameters (window type and length, overlap percentage), neighborhood size $\ing{T}$, spectral contents of the excitation signal and noise, and, naturally, room acoustics. \revis{We will see later that the dominant wavefront is positioned at the zero-delay index of the RdRIR representation, \emph{i.e.}, in the middle of the time series. Hence, when the dominant wavefront is due to direct sound, capturing the early echoes requires the STFT frame length to be at least twice the mixing time value $\bar{\tau}_{\max}$. The estimator presents certain advantages and drawbacks. Conveniently, it requires only the information about the activity of the target sound source. This is also related to its susceptibility to directional interference, thus, a reliable voice activity detector (VAD) is implied. While it can adapt to changes in the acoustic environment, its performance tend to degrade in highly dynamic scenarios. On the other hand, if more refined information is available, such as the noise covariance matrix, one could conceive adaptations of other well-known RTF estimation techniques, \emph{e.g.}, the covariance subtraction and covariance whitening methods \cite{markovich2018performance,jarrett2017theory}. }

\begin{algorithm}
\caption{Self-steering GTVV estimator at a frame $\ing{t}_0$}\label{algGTVV}
\revis{\begin{algorithmic}
\Require STFT tensor $\{ \hat{\vect{b}}(\ing{f},\ing{t}) \mid \ing{t}\in \ing{t_0} + [\ing{-T/2},\ing{T/2}]$, $\ing{f} \in [0, \ing{K}), \; \ing{l} \in[0,(\ing{L}+1)^2)$\}, parametric SH dictionary $\{\vect{y}(\theta,\phi)\}_{(\theta,\phi)}, \; \ing{num\_iter}$
\Compute $\hat{\phi}_{\ing{l}}^2(\ing{f,t})=|\hat{b}_{\ing{l}}(\ing{f,t})|^2$ and $\hat{\phi}_{\ing{l}',\ing{l}}(\ing{f,t}) = \hat{b}_{\ing{l}'}(\ing{f,t})\hat{b}^*_{\ing{l}}(\ing{f,t})$
\Assemble $\hat{\vect{\phi}}_{\ing{l}}^2(\ing{f})$ and $\hat{\mtrx{\Phi}}_{:,\ing{l}}(\ing{f})$ from eq. \eqref{eqPhiVect} and \eqref{eqPhiMtx}
\State Set $\vect{w} = \transp{\left[ \begin{smallmatrix} 1 & 0 & 0 & \hdots & 0 \end{smallmatrix} \right]}$
\For{$\ing{iter} \in [1,\ing{num\_iter}]$}
    \For{$\ing{l} \in [0,(\ing{L}+1)^2-1]$}
    \State $\hat{\vect{v}}(\ing{f}) \gets$ solve \eqref{eqCompactApproxVect} for each $\ing{f}$
    \State $\vect{v}(\tilde{t}) \gets \mathcal{F}^{-1} \left(\hat{\vect{v}}(\ing{f}) \right)$
    \EndFor
    \State $(\theta_0,\phi_0) \gets \argmax_{(\theta,\phi)} \transp{\vect{v}(\tilde{t}=0)}\vect{y}(\theta,\phi)$
    \State $\vect{w} \gets \vect{y}(\theta_0,\phi_0)/(\ing{L}+1)^2$
\EndFor
\Myreturn $\vect{v}(\tilde{t})$, $\hat{\vect{v}}(\ing{f})$, $(\theta_0,\phi_0)$
\end{algorithmic}}
\end{algorithm}

In addition to the previously discussed uncertainties, another unknown parameter is the DoA of the source of interest, which is often a required parameter to design the beamformer vector $\hat{\vect{w}}(f)$. The following subsection is dedicated to the derivation of a closed-form expression of GTVV, where we demonstrate that -- under certain conditions -- GTVV can be used to directly infer the source's DoA. However, we have observed that even if these conditions do not hold, the GTVV vector $\vect{v}(t=0)$ is usually a good approximation of the SH encoding vector parametrized by a steering angle pointed in the vicinity of DoA. We exploit this observation to devise a heuristic scheme that improves the DoA estimate iteratively. Indeed, a ``well-behaved'' GTVV representation maintains $\vect{v}(t=0)$ that is invariant to the slight changes in $\hat{\vect{w}}(f)$, \emph{i.e.}, it should always point towards DoA. We, therefore, use the current DoA estimate to (re-)steer the beamformer and compute a new GTVV representation.  Starting with the omnidirectional reference (the classical relative IR), we monitor the difference in DoA between iterations to deduce whether GTVV has ``converged''. Typically, this procedure requires no more than ten iterations. \revis{For reader's convenience, its pseudocode is given in Alg.~\ref{algGTVV} (the specification of the applied signal-independent beamformer is given in Section~\ref{secExperiments}).}

\subsection{Closed-form expression}

In order to derive a closed-form expression of GTVV, we will treat the numerator and denominator of the GFVV expressions \eqref{eqGFVVcompact} separately. 

The numerator $\hat{\vect{h}}(f)$ is simply the ``early'' part of ATF, normalized by the amplitude of direct component $\hat{a}_0(f)e^{-j2\pi f \bar{\tau}_0 }$. Hence, its time domain counterpart is the early part of RIR, normalized and shifted to temporal origin - dubbed hereafter \emph{Reduced Room Impulse Response (RdRIR)}:
\begin{equation}\label{eqRdRIR}
   \vect{h}(t) = \delta(t)\vect{y}_0 + \sum\limits_{\ing{n}=1}^{\ing{N}-1} g_{\ing{n}}(t-\tau_{\ing{n}})\vect{y}_{\ing{n}}.
\end{equation} 
Note that, under the assumptions that have been stated earlier, $\hat{g}_{\ing{n}}(f)$ is a real-valued, nonnegative and even function of frequency. Hence, it is easy to show that $g_{\ing{n}}(t) = \mathcal{F}^{-1}\left(\hat{g}_{\ing{n}}(f)\right)$ is also real and even, and that it attains global maximum at $t=0$. Since we also assumed that $g_{\ing{n}}(t)$ has compact support, temporally well-separated wavefronts (having sufficiently distinct delays $\tau_{\ing{n}}$), could be identified by observing peaks of its delay-magnitude time series $\zeta_{\vect{h}}(t)$:
\begin{equation}\label{eqDelMagn}
    \zeta_{\vect{h}}(t) = \norm{\vect{h}(t)}{2}.
\end{equation}

\begin{figure}
  \includegraphics[width=\columnwidth]{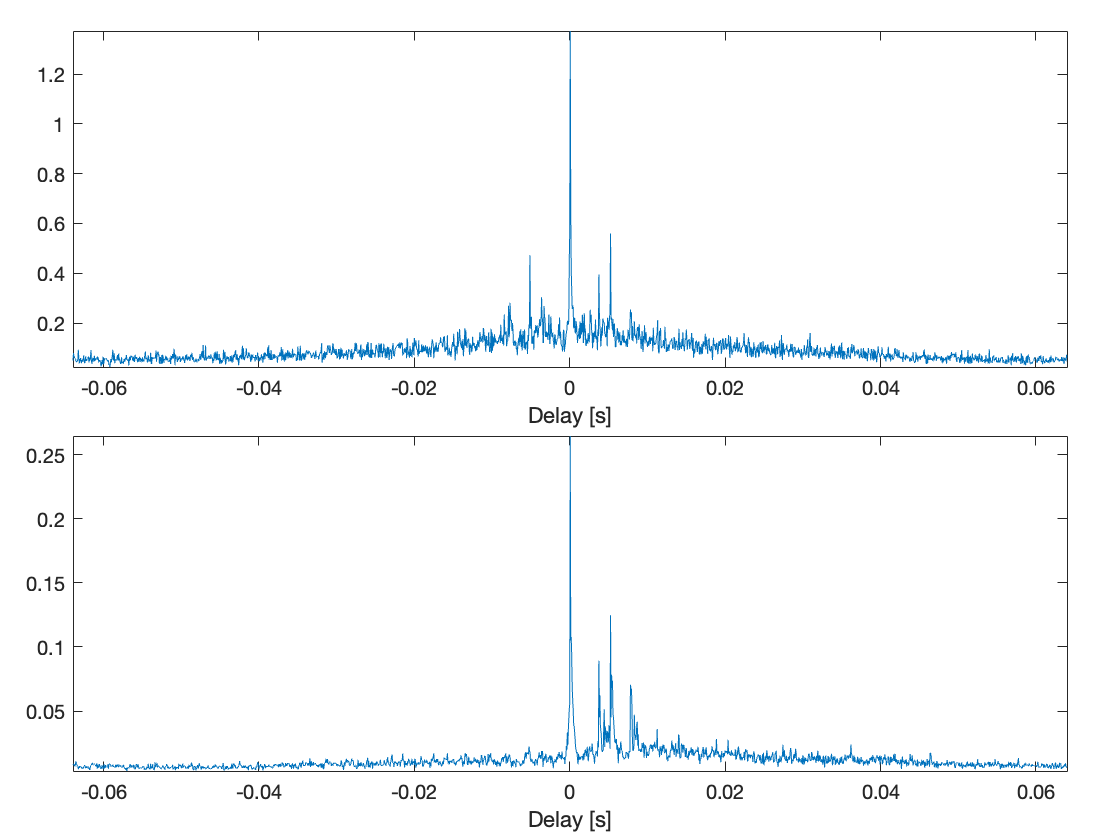}
  \caption{The GTVV estimate's delay-magnitude $\zeta_{\vect{v}}(t)$ without (top) and with (bottom) condition \eqref{eqTaylorCondition} satisfied. The latter is approximately causal, as predicted.}\label{figTaylor}
\end{figure}

An explicit solution of $a^{-1}(t)$ requires more attention. We introduce an additional hypothesis: 
\begin{equation}\label{eqTaylorCondition}
  \left| \sum\limits_{\ing{n}=1}^{\ing{N}-1} \hat{\kappa}_{\ing{n}}(f) e^{-j2\pi f \tau_{n} } \right| < 1, \; \forall f,
\end{equation}
which is also a sufficient condition for assuring that the impulse response of the reference is a minimum-phase filter \cite{cassioli2009minimum}. The importance of this condition for the extraction of wavefront parameters will be discussed in the remainder of this section. For the time being, we motivate its interest by visually inspecting two instances of the time series $\zeta_{\vect{v}}(t) = \norm{\vect{v}(t)}{2}$ in Fig.~\ref{figTaylor}, with and without condition \eqref{eqTaylorCondition} satisfied by the vector-valued GTVV function $\vect{v}(t)$. 

The requirement \eqref{eqTaylorCondition} is obviously granted if the magnitude of the direct component is larger than the cumulative magnitude of all reflections in the reference signal: 
\begin{equation}\label{eqTaylorConditionStronger}
  \sum\limits_{\ing{n}=1}^{\ing{N}-1} \left|\hat{\kappa}_{\ing{n}}(f) \right| = \sum\limits_{\ing{n}=1}^{\ing{N}-1} \hat{g}_{\ing{n}}(f) \left|\hat{\beta}_{\ing{n}}(f) \right|  < 1, \; \forall f.
\end{equation}

Clearly, this is highly unlikely if the reference is the omnidirectional channel. Instead, by using a beamformer steered towards DoA, this hypothesis becomes more and more plausible with the increase in Ambisonic order. Indeed, for beamformers such as maximum-directivity or Minimum Variance Distortionless Response (MVDR) \cite{jarrett2017theory}, having $|\htransp{\hat{\vect{w}}(f)}\vect{y}_n| < \beta_0$ leads to $\lim_{\ing{L}\to\infty} |\hat{\beta}_{\ing{n}}(f)| = 0$, due to the completeness property of spherical harmonics \cite{rafaely2015fundamentals}. Intuitively, with the increase in $\ing{L}$, the spatial response of the beamformer approaches the delta function centered around DoA \cite{jarrett2017theory}. Alternatively, one may consider forcing spatial nulls in the directions of strong reflectors (known a priori), \emph{e.g.} by using the Linearly Constrained Minimum Variance (LCMV) \cite{gannot2017consolidated} beamformer.

Should the condition \eqref{eqTaylorCondition} hold, we can reformulate the GFVV denominator $\hat{a}^{-1}(f)$ in \eqref{eqGFVVcompact} through the Taylor (geometric) series expansion. In the following, we omit the frequency variable $f$ for brevity, and let $\gamma_{\ing{n}} := -\hat{\kappa}_{n}(f) e^{-j2\pi f \tau_{n} }$. Then, the denominator in \eqref{eqFDVVinst} becomes
\begin{equation}
  \frac{1}{\hat{a}(f)} = \frac{1}{1 - \sum\limits_{\ing{\ing{n}}=1}^{\ing{N}-1} \gamma_{\ing{n}}} = \sum\limits_{\ing{k}=0}^{\infty} \left(-\sum\limits_{\ing{n}=1}^{\ing{N}-1} \gamma_{\ing{n}} \right)^{\ing{k}} := \sum\limits_{\ing{k}=0}^{\infty} \lambda_{\ing{k}},
\end{equation}
where each element $\lambda_{\ing{k}}$ of the last sum is developed using multinomial theorem into
\begin{equation*}
   \lambda_{\ing{k}} = \sum_{\ing{i}_1 + \ing{i}_2 + \hdots + \ing{i}_{\ing{N}-1} = \ing{k}} \frac{\ing{k}!}{\ing{i}_1!\ing{i}_2!\hdots \ing{i}_{\ing{N}-1}!} \prod\limits_{\ing{q}=1}^{\ing{N}-1} \gamma_{\ing{q}}^{\ing{i}_{\ing{q}}}.
\end{equation*}

Evaluating $\lambda_{\ing{k}}$ for $\ing{k} = 0, 1, 2, 3 \hdots$ yields:
\begin{align}
    & \lambda_0 = 1, \nonumber\\ 
    & \lambda_1 = \sum\limits_{\ing{i}} \gamma_{\ing{i}}, \nonumber \\
    &\lambda_2 = \sum\limits_{\ing{i}}  \gamma_{\ing{i}}^2 + 2\sum\limits_{\ing{i}} \sum\limits_{\ing{m} \neq \ing{i}} \gamma_{\ing{i}}\gamma_{\ing{m}}, \\
    & \lambda_3 = \sum\limits_{\ing{i}} \gamma_{\ing{i}}^3 + 3\sum\limits_{\ing{i}} \sum\limits_{\ing{m}\neq \ing{i}} \gamma_{\ing{i}}^2 \gamma_{\ing{m}}  
      + 6 \sum\limits_{\ing{i}} \sum\limits_{\ing{m}\neq \ing{i}} \sum\limits_{\ing{p}\neq \{\ing{i}, \ing{m}\}} \gamma_{\ing{i}}\gamma_{\ing{m}}\gamma_{\ing{p}} \nonumber \\
    & \hdots \nonumber
\end{align}
where all sums correspond to indices in $[\ing{1,N}]$. Due to directional beamforming, we expect only a subset of reflections to have non-negligible magnitudes $|\kappa_{\ing{n}}(f)| \gg 0$, $\ing{n} \in [1,\ing{N}-1]$ (with the size of this subset decreasing with the increase in Ambisonic order, as discussed before). Therefore, the magnitudes of ``cross-terms'' in the expressions above are more likely to diminish \revis{than} the leftmost terms that correspond to isolated reflections. We simplify the expression by aggregating all cross terms in a single variable $\eta(f)$, hence the GFVV denominator is represented as
\begin{equation}
  \sum\limits_{\ing{k}=0}^{\infty} \lambda_{\ing{k}} = 1 + \sum_{\ing{i}=1}^{\ing{N}-1}\sum\limits_{\ing{k}=1}^{\infty} \left(-\hat{\kappa}_{\ing{i}}(f) \right)^{\ing{k}}e^{-j2\pi f \ing{k} \tau_{\ing{i}}} + \hat{\eta}(f).
\end{equation}

In time domain, this yields the following expression:
\begin{align}\label{eqDenominator}
  a^{-1}(t) & = \delta(t) + \sum_{\ing{i}=1}^{\ing{N}-1}\sum\limits_{\ing{k}=1}^{\infty} (-1)^{\ing{k}}\kappa_{\ing{i}}^{\ast \ing{k}}(t- \ing{k}\tau_{\ing{i}}) + \eta(t), \\
  \text{where} \; \kappa_{\ing{i}}^{\ast \ing{k}}(t) = & \mathcal{F}^{-1} \left( \hat{\kappa}_{\ing{i}}(f)^{\ing{k}} \right) = \overbrace{(\kappa_{\ing{i}} \ast \kappa_{\ing{i}} \ast \hdots \ast \kappa_{\ing{i}})}^{\ing{k}-1 \; \text{convolutions}} (t).\nonumber
\end{align}
Plugging \eqref{eqRdRIR} and \eqref{eqDenominator} into \eqref{eqGTVVinst}, and manipulating the terms within, produces
\begin{multline}\label{eqGTVVcf}
  \vect{v}(t) = \delta(t)\vect{y}_0 + \sum\limits_{\ing{n}=1}^{\ing{N}-1}\sum\limits_{\ing{k}=1}^{\infty} (-1)^{\ing{k}} \kappa_{\ing{n}}^{\ast \ing{k}}(t - \ing{k}\tau_{\ing{n}}) \ast \\
   \left(\vect{y}_0 \delta(t) - \vect{y}_{\ing{n}}\beta_{\ing{n}}^{-1}(t) \right) + \tilde{\eta}(t),
\end{multline}
where $\tilde{\eta}(t)$ again accounts for $\eta(t)$, augmented by additional cross-convolutions between different reflections. One can make several observations of the GTVV representation \eqref{eqGTVVcf}. First, due to the assumed compact support of $\kappa_{\ing{n}}(t)$, GTVV is approximately causal, \emph{i.e.}, $\vect{v}(t<0) \approx \vect{0}$. Second, if the stronger condition \eqref{eqTaylorConditionStronger} holds, we expect GTVV to be somewhat sparse (as the energy of $\kappa_{\ing{n}}^{\ast \ing{k}}(t)$ decreases with $\ing{k}$). Third, interestingly, GTVV still allows us to immediately identify the SH vector corresponding to direct component, by evaluating $\vect{v}(t=0)$, as for RdRIR in \eqref{eqRdRIR}. However, even by neglecting the cross-terms $\tilde{\eta}(t)$, it is obvious that the remainder of the GTVV expression is more complex, presenting itself as a series of repeated convolutions with alternating sign, for each wavefront $\ing{n}$. 

Hence, without additional assumptions, we cannot easily identify the remaining wavefronts. For instance, if we conjecture that the initial terms ($\ing{k}=1$) of each series are not strongly affected by another series \emph{and} that $\beta_{\ing{n}}(0) \ll 1$, then the largest peaks of $\zeta_{\vect{v}}(t)$ would likely\footnote{It may still happen that later terms ($\ing{k}>1$) of dominant reflections have larger peaks than the initial terms of weaker wavefronts!} correspond to SH vectors $\vect{y}_{\ing{n}}$, \emph{i.e.}
\begin{equation}
  \vect{v}(\tau_{\ing{n}}) = g_{\ing{n}}(0)\left(\vect{y}_{\ing{n}} - \beta_{\ing{n}}(0)\vect{y}_0 \right) \approx g_{\ing{n}}(0)\vect{y}_{\ing{n}}.
\end{equation}

If $\beta_{\ing{n}}(0) \ll 1$ does not hold, yet we still assume that $\vect{v}(\tau_{\ing{n}})$ could be isolated (\emph{e.g.}, for the strong reflections), we can exploit the fact that $\vect{y}_0$ can be pre-estimated from $\vect{v}(t=0)$ to estimate the wavefront vector $\vect{y}_{\ing{n}}$. In this case, we consider only a wideband beamformer $\vect{w}$ and propose to solve a nonlinear optimization problem:
\begin{equation}\label{eqCost}
  (\theta_{\ing{n}},\phi_{\ing{n}}) = \argmin_{(\theta,\phi)} \transp{\vect{v}(\tau_{\ing{n}})}\left(\mtrx{I} - \vect{y}_0 \transp{\vect{w}}  \right) \vect{y}(\theta,\phi),
\end{equation}
where $\vect{y}(\theta,\phi)$ is a SH vector for the given azimuth and elevation parameters. In the cost function \eqref{eqCost}, we have used the expression for the (constant) spatial response of a wideband beamformer $\beta_{\ing{n}} = \transp{\vect{w}}\vect{y}_{\ing{n}}$. To avoid explicitly solving the optimization problem above, one may use a dictionary of normalized SH encoding vectors $\vect{y}(\theta,\phi)$, parametrized from a discrete grid of directions $\{ (\theta,\phi) \}$, and choose the atom most correlated with $\transp{\vect{v}(\tau_{\ing{n}})}\left(\mtrx{I} - \vect{y}_0 \transp{\vect{w}}  \right)$. Unfortunately, the matrix $\mtrx{I} - \vect{y}_0 \transp{\vect{w}}$ is not invertible (otherwise, one could directly obtain an estimate of $\vect{y}_{\ing{n}}$), which is easy to show by applying the Sherman-Morrison formula \cite{golub2013matrix}.

Finally, we remark that the presented derivation does not directly depend on the DoA direction $\vect{y}_0$ used for the distortionless constraint $\htransp{\vect{w}(f)}\vect{y} = \text{const}$, as long as the Taylor series condition \eqref{eqTaylorCondition} holds. In other words, if a sufficiently selective beamformer is focused on some other wavefront $\ing{r}$ (\emph{e.g.}, a dominant reflection), one would replace $\vect{y}_0$ by $\vect{y}_{\ing{r}}$, and a similar expression applies:
\begin{multline}\label{eqGTVVcf_refl}
  \vect{v}(t) = \delta(t)\vect{y}_{\ing{r}} + \sum\limits_{\ing{n}\neq\ing{r}}^{\ing{N}-1}\sum\limits_{\ing{k}=1}^{\infty} (-1)^{\ing{k}} \kappa_{\ing{n}}^{\ast \ing{k}}(t - \ing{k}\tau_{\ing{n}}) \ast \\
   \left(\vect{y}_{\ing{r}} \delta(t) - \vect{y}_{\ing{n}}\beta_{\ing{n}}^{-1}(t) \right) + \tilde{\eta}(t),
\end{multline}
except that the quantities $g_{\ing{n}}$ and $\tau_{\ing{n}}$ are now relative to the absolute gain $a_{\ing{r}}(f)$ and ToA $\bar{\tau}_{\ing{r}}$ of this wavefront, respectively. As a consequence, relative gains $g_{\ing{n}}$ would not be bounded by $1$, relative delays $\tau_{\ing{n}} = \bar{\tau}_{\ing{n}} - \bar{\tau}_{r}$ could have negative values, and $\vect{v}(t=0)$ would encode the reflection direction $(\theta_{\ing{r}},\phi_{\ing{r}})$. Compared to the GTVV computed using DoA, this variant would be ``shifted'' to the left by $|\tau_0|$.

\section{Estimation of reduced RIR}
\label{secRdRIR}

We have argued that GTVV is better adapted to reverberant acoustic conditions than the ``standard'' relative impulse response for which the reference signal is the zero-order Ambisonic channel. Nevertheless, it is still limited by the spatial selectivity of the applied beamformer - for example, if the signal-independent maximum directivity beamformer is used, its directivity will be proportional to the square of Ambisonic order \cite{jarrett2017theory}. However, affordable Ambisonic microphone arrays usually do not provide very high order Ambisonic formats - most often, they are only capable of recording the FOA signals \cite{lee2021multichannel}. Furthermore, the frequency support of higher order channels progressively decreases with the HOA order, as noise amplification at low frequencies, and spatial aliasing at high frequencies start to kick-in \cite{daniel2004further}. Unfortunately, the favorable theoretical properties of GTVV tend to diminish at low Ambisonic orders, due to the inability of the applied beamformer to effectively suppress the reflections. The problem is further exacerbated with the increase in the microphone-to-source distance, since more reflections fall within the main lobe of the beamformer. In practice, we observe that the GTVV imprint is no longer causal (as seen in Fig.~\ref{figTaylor}), and that the estimated directions are less accurate. 

Moreover, even when the GTVV expression \eqref{eqGTVVcf} remains valid, identifying the directions and delays by peak-picking is not straightforward, as discussed in the previous section. In fact, such a ``well-behaved'' GTVV can be seen as the reduced RIR \eqref{eqRdRIR}, convolved by the minimum-phase filter \eqref{eqDenominator}. The consequence is that the same reflection is infinitely “echoed” at the time instances corresponding to integer multiples of its relative delay, with the alternating sign and the decreasing magnitude. Thus, these series can interfere with one another, altering the information within, or even masking the presence of weaker reflections. Undoubtedly, it is much easier and intuitive to extract information directly from RdRIR \eqref{eqRdRIR}. In this section, we propose a simple method to estimate the latter from the observed GTVV time series, even if the convergence condition \eqref{eqTaylorCondition} is not satisfied. The development is based on the celebrated Pad\'e-Prony method for the pole-zero modeling \cite{hayes2009statistical}, which is very similar to traditional Autoregressive Moving Average (ARMA) model for stochastic time series. 

In the following, we consider a beamformer steered towards DoA, since the same method could be straightforwardly adapted when other wavefronts are considered. We start by rewriting \eqref{eqGTVVinst} as 
\begin{equation}\label{eqRdRIRconv}
  \left(\vect{v} \ast a\right)(t) = \vect{h}(t),
\end{equation}
and recall that $a(t) = \mathcal{F}^{-1}\left( \htransp{\vect{w}(f)}\vect{h}(f) \right)$, \emph{i.e.},

\begin{align}
  a(t) &= \mathcal{F}^{-1}\left( 1 + \sum\limits_{\ing{n}=1}^{\ing{N}-1} \hat{\kappa}_{\ing{n}}(f) e^{-j2\pi f \tau_{\ing{n}} } \right) \nonumber \\
  & = \delta(t) + \sum\limits_{\ing{n}=1}^{\ing{N}-1} \kappa_{\ing{n}}(t - \tau_{\ing{n}}).\label{eqFilter_a}
\end{align}

Since we have assumed that all $\kappa_{\ing{n}}(t)$ have compact support, $a(t)$ is a causal filter (but, not necessarily a minimum-phase!). We already know from \eqref{eqRdRIR} that RdRIR $\vect{h}(t)$ is a causal vector sequence, \emph{i.e.}, $\vect{h}(t<0) \approx \vect{0}$. Moreover, RdRIR has finite support - beyond the relative delay $\tau_{\max} = \bar{\tau}_{\max} - \bar{\tau}_0$, \emph{i.e.}, relative to the mixing time $\bar{\tau}_{\max}$, we expect $\vect{h}(t > \tau_{\max}) \approx \vect{0}$ to hold as well (being the feature of the ``denoising'' estimator presented in subsection~\ref{ssec:estimation}). Likewise, one may argue that the early part of RIR (hence, RdRIR) is relatively sparse \cite{gannot2017consolidated,xue2016cross,xue2017frequency}. The filter $a(t)$ would be even sparser, as we expect the beamforming operation to suppress certain reflections in the reference signal. Finally, one may observe from \eqref{eqRdRIR} that, for any $t$, the zero-order entry of $\vect{h}(t)$ is non-negative. Our aim is to take advantage of all this prior knowledge to estimate $a(t)$ directly from $\vect{v}(t)$, and then extract $\vect{h}(t)$ by convolving its estimate with GTVV, as in \eqref{eqRdRIRconv}. 

As usual in digital signal processing, we do not handle continuous functions $\vect{v}(t)$, $\vect{h}(t)$ and $a(t)$, but their sampled versions. We make a leap of faith and assume that the latter are not substantially affected by aliasing, meaning that the properties discussed above are generally preserved. We denote by $\ing{j}$ the time sample index taking values\footnote{We intentionally permit negative indexing, to preserve the intuition that the temporal dimension is centered at zero.} in $[-\ing{J}/2+1,\ing{J}/2]$, within an STFT frame of length ${\ing{J} \in 2\mathbb{N}}$. Thus, both GTVV and RdRIR are represented by the real-valued matrices $\mtrx{V}$ and $\mtrx{H}$ of size $(\ing{L}+1)^2 \times \ing{J}$, \emph{i.e.} their columns $\vect{v}_{:,\ing{j}}$ (accordingly, $\vect{h}_{:,\ing{j}}$) are akin to evaluating $\vect{v}(t)$ and $\vect{h}(t)$ at some time instant $t$. Analogously, the rows $\vect{v}_{\ing{l},:}$ (accordingly, $\vect{h}_{\ing{l},:}$) correspond to the $\ing{l}$\textsuperscript{th} channels of the two representations. The filter $a(t)$ is replaced by a vector $\vect{a}\Rset{\ing{j}_{\max}+1}{1}$, where the hyperparameter $\ing{j}_{\max}$ denotes the sample index corresponding to $\tau_{\max}$, \emph{i.e.}, the assumed relative delay of the ``last'' wavefront in RdRIR. The coefficients $a_{\ing{j}<0}$ and $a_{\ing{j}>\ing{j}_{\max}}$ are assumed to be zero, hence, these are not included in the estimation vector $\vect{a}$. 


Setting aside the sparsity hypothesis for now, note that enforcing $\zeta_{\vect{h}}(t)=0$, for $t<0$ and $t>\tau_{\max}$, amounts to minimizing the following cost function: 
\begin{equation}\label{eqOptim_a}
  \min_{\vect{a}} \sum\limits_{\ing{l}=0}^{(\ing{L}+1)^2-1} \sum\limits_{\ing{j} \notin [0,\ing{j}_{\max}]} (v_{\ing{l},:} \ast a)^2_{\ing{j}}, \; \; \text{s.t.} \; \; a_0=1.
\end{equation}
The equality constraint is due to $a(0)=\delta(0)$ in \eqref{eqFilter_a}, with the Dirac delta distribution replaced by the Kronecker delta function in the discrete version. We remark that \eqref{eqOptim_a} is a particular multichannel linear prediction problem, with the filter $\vect{a}$ being common for all channels $\ing{l}\in [0, (\ing{L+1})^2-1]$. This is advantageous - since the problem is overdetermined, the estimate of $\vect{a}$ should be more resilient to GTVV estimation errors and noise. Furthermore, the estimation should become more accurate as the channel order $\ing{L}$ increases. 



There are multiple approaches of addressing linear prediction problems, but probably the most well-known are the autocorrelation method and the covariance method \cite{hayes2009statistical}. In both cases, solving the constrained quadratic problem comes down to a linear system, compactly written as
\begin{align}
  & \sum_{\ing{j}=1}^{\ing{j}_{\max}} a_{\ing{j}} r(\ing{j},\ing{s}) = -r(0,\ing{s}), \label{eqLPsolution}\\
  \text{where} \; & r(\ing{j},\ing{s}) = \sum\limits_{\ing{l}=0}^{(\ing{L+1})^2-1} r_{\ing{l}}(\ing{j},\ing{s}) =  \sum\limits_{\ing{l}=0}^{(\ing{L+1})^2-1} \sum_{\ing{j}'} v_{\ing{l}, \ing{j'-j}} v_{\ing{l},\ing{j'-s}}.\label{eqLPcoeffs}
\end{align}

The two methods differ in the way they deal with the signal edges, \emph{i.e.}, how they define the range of the summation variable $\ing{j}'$. The autocorrelation method applies zero-padding ($v_{\ing{l},\ing{j}'} =0$, for $\ing{j}' \in [0,\ing{j}_{\max}]$ and $\ing{j}' \notin [-\ing{J}/2+1,\ing{J}/2]$), while the covariance method considers only valid parts of the convolution (where the two sequences overlap and $\ing{j'} \notin [0,\ing{j}_{\max}]$), and discards the rest. Therefore, the coefficients $r(\ing{j},\ing{s})$ would be somewhat different, yielding different solutions. \revis{Particularly, the filter $\vect{a}$ estimated by the autocorrelation method is always minimum-phase \cite{hayes2009statistical}, but the corresponding linear system has Toeplitz structure, hence it can be solved by the Levinson-Durbin algorithm \cite{golub2013matrix} with $O((\ing{j}_{\max}+1)^2)$ time complexity. This is significantly more efficient compared to $O((\ing{j}_{\max}+1)^3)$ of the covariance method.} 

\revis{Furthermore, calculating the coefficients \eqref{eqLPcoeffs} of the normal equations \eqref{eqLPsolution} generally requires $O((\ing{L}+1)^2\ing{J}^2 )$ multiplications, but for the autocorrelation method this cost is reduced, thanks to the duality of autocorrelation and power spectrum \cite{hayes2009statistical}. Due to the symmetry property of autocorrelation, we have $r_{\ing{l}}(\ing{j},\ing{s}) = r_{\ing{l}}(\ing{j}-\ing{s}) = r_{\ing{l}}(\ing{s}-\ing{j})$. Now define 
\begin{align}
    v^{-}_{\ing{l}, \ing{j}} &= 
    \begin{cases}
        v_{\ing{l}, \ing{j}}, & \ing{j} \in [-\ing{J}/2-1,0), \\
        0, & \text{otherwise,}
    \end{cases} \label{eqVminus} \\
    v^{+}_{\ing{l}, \ing{j}} &= 
    \begin{cases}
        v_{\ing{l}, \ing{j}}, & \ing{j} \in (\ing{j}_{\max},\ing{J}/2], \\
        0, & \text{otherwise,}
    \end{cases} \label{eqVplus}
\end{align}
}
\revis{and let
\begin{align}
    r_{\ing{l}}^{-}(\ing{j}-\ing{s}) &= \sum\limits_{\ing{j}'={-\ing{J}/2-1}}^{\ing{J}/2} v^-_{\ing{l}, \ing{j'-j}} v^-_{\ing{l},\ing{j'-s}}=\mathcal{F}^{-1} \left(| \hat{\vect{v}}^{-}_{\ing{l},:}|^2 \right)_{\ing{j}-\ing{s}}, \label{eqRminus} \\
    r_{\ing{l}}^{+}(\ing{j}-\ing{s}) &= \sum\limits_{\ing{j}'={-\ing{J}/2-1}}^{\ing{J}/2} v^+_{\ing{l}, \ing{j'-j}} v^+_{\ing{l},\ing{j'-s}} = \mathcal{F}^{-1} \left( | \hat{\vect{v}}^{+}_{\ing{l},:} |^2 \right)_{\ing{j}-\ing{s}} \label{eqRplus},
\end{align}
where $\hat{\vect{v}}^{-}_{\ing{l},:}$ and $\hat{\vect{v}}^{+}_{\ing{l},:}$ are the frequency representations of $\vect{v}^{-}_{\ing{l}, :}$ and $\vect{v}^{+}_{\ing{l}, :}$, respectively. Having $r_{\ing{l}}(\ing{j}-\ing{s}) = r_{\ing{l}}^{-}(\ing{j}-\ing{s}) + r_{\ing{l}}^{+}(\ing{j}-\ing{s})$ and 
\begin{multline}\label{eqACcoeffs}
    r(\ing{j}, \ing{s}) = \sum\limits_{\ing{l}=0}^{(\ing{L+1})^2-1} r_{\ing{l}}(\ing{j} -\ing{s}) \\= \mathcal{F}^{-1} \left( \sum\limits_{\ing{l}=0}^{(\ing{L+1})^2-1} (| \hat{\vect{v}}^{-}_{\ing{l},:}|^2 + | \hat{\vect{v}}^{+}_{\ing{l},:}|^2) \right)_{\ing{j}-\ing{s}}, 
\end{multline}
we obtain the multichannel linear prediction coefficients at $O((\ing{L+1})^2\ing{J}\log{\ing{J}} )$ computational cost.
}

We note that the presented approach is a variant of classical Prony-like estimation, lauded for its computational efficiency, yet a more elaborate technique may be applied. For instance, one could perform alternating minimization to improve the estimates of $a(t)$ and $\vect{h}(t)$ iteratively, in the spirit of the Steiglitz-McBride algorithm \cite{steiglitz1965technique}. Therefore, to incorporate the sparsity and non-negativity assumptions, we propose to jointly optimize the two variables:
\begin{align}\label{EQOPTIM_AH}
  \min_{\vect{a},\mtrx{H}}  & \sum_{\ing{j}} \norm{\vect{h}_{:,\ing{j}}}{2} \nonumber \\ 
  \text{s.t.} \; & \vect{h}_{\ing{l},:} = \vect{v}_{\ing{l},:} \ast \vect{a}, \\
  & h_{\forall \ing{l},\ing{j}<0} = h_{\forall \ing{l},\ing{j}>\ing{j}_{\max}}=0, \nonumber \\
  & h_{0,\forall \ing{j}} \geq 0 \; \text{and} \; a_0 = 1. \nonumber
\end{align}

\begin{figure}
  \includegraphics[width=\columnwidth]{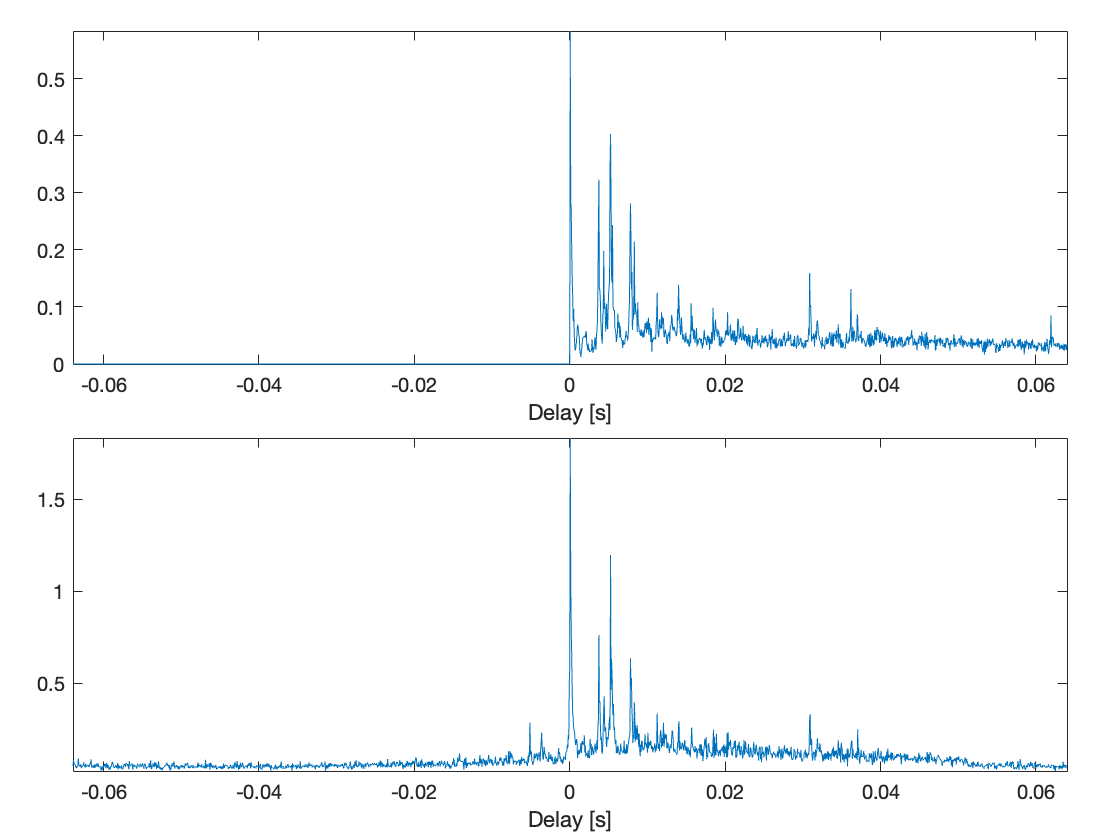}
  \caption{Delay-magnitude $\zeta_{\vect{h}}(t)$ of the ground truth (top) and estimated (bottom) RdRIR, recovered from the acausal GTVV representation given in Fig~\ref{figTaylor} (top).}\label{figExample}
\end{figure}

However, the imposed modeling constraints make this problem inconsistent in practice (\emph{e.g.} due to estimation errors and noise). While one could reformulate the problem such that the constraints are relaxed - for instance, by introducing a squared norm penalty instead of the first equality constraint, it would require introducing a new regularization hyperparameter. Instead, we propose using Alternating Directions Method of Multipliers (ADMM), a first-order optimization framework based on Douglas-Rachford splitting \cite{eckstein1992douglas}. ADMM is particularly effective for optimization problems involving linearly dependent variables, provided that the solutions of intermediate optimization problems are efficiently obtained. Another convenient feature of ADMM is that, in the inconsistent setting, its iterates could produce the best approximation pair, \emph{i.e.}, a pair of estimates of $\vect{a}$ and $\mtrx{H}$ for which the residuals $\vect{h}_{\ing{l},:} - \vect{v}_{\ing{l},:} \ast \vect{a}$ attain the lowest norm \cite{aragon2020douglas}. In Appendix~\ref{appADMM}, we instantiate ADMM for the problem \eqref{EQOPTIM_AH} - an interested reader can easily derive the algorithm by following the tutorial article \cite{boyd2011distributed} by Boyd et al. Fig.~\ref{figExample} illustrates an example of the reconstructed RdRIR using the proposed ADMM.

\section{Experiments}
\label{secExperiments}

We evaluate RdRIR estimation on data generated using simulated and recorded Ambisonic RIRs. In particular, the autocorrelation (AC), covariance (COV) and ADMM methods applied to GTVV are benchmarked. \revis{As baselines, we use the canonical MCLMS algorithm, as well as the noise-robust multichannel frequency-domain least mean squares (RNMCFLMS) \cite{haque2008noise} version. Both least mean squares (LMS) implementations are from the \emph{Blind System Identification and Equalization (BSIE)} toolbox \cite{BSIEtoolbox}. As additional baselines, we use the ``plain'' GTVV and TDVV representations. }

The "fully blind" scenario is assured in all experiments, \emph{i.e.}, the algorithms only have access to the (noisy) observed signals, which are generated by convolving the multichannel RIRs with $10$~s of speech data, \revis{taken from the publicly available LibriSpeech corpus \cite{panayotov2015librispeech}}. \revis{The TDVV and GTVV representations are estimated from these measurements using the approach described in \ref{ssec:estimation}, with the latter being obtained through the ``self-steering'' heuristics, explained at the end of the same subsection.} The number of frames $\ing{T}$ used for the estimation corresponds to the $0.5$~s buffer. \revis{The GTVV reference signal is obtained using a ''regular'' (or "Plane-Wave Decomposition" or "Maximum-Directivity") beamformer \cite{jarrett2017theory} pointing towards the iteratively estimated main DoA $(\tilde{\theta_0},\tilde{\phi_0})$, \emph{i.e.}: $\vect{w} = \vect{y}(\tilde{\theta_0},\tilde{\phi_0}) / (\ing{L}+1)^2 $, 
assuming that the spherical harmonic function basis is 3D-Normalized \cite{daniel2003nfchoa}.} Three HOA orders are considered: $\ing{L}\in\{1,2,3\}$, while the common sampling rate is set to $f_s=16$~kHz. The STFT representation is computed by applying Tukey window of length $0.128$~s, with $75\%$ overlap between succeeding frames. The Hendriks algorithm \cite{gerkmann2011unbiased} is used as VAD. \revis{When applied to the \emph{clean} speech data, this algorithm estimates that about $50\%$ of all STFT frames are voiced.}

Since the RdRIR representation is invariant to global scale and ToA offset, it cannot be directly compared to the ground truth multichannel RIRs. Moreover, since the measured RIRs are sampled versions of continuous impulse responses, some of their segments may have undergone sign inversion (due to the action of the anti-aliasing filter). Assuming the same filter has been applied to all channels, it suffices to observe the sign of the omnidirectional component, and then, if the latter is negative, change the signs of all channels at the given sample. Following the sign correction, RIRs are shifted to temporal origin, such that the strongest wavefront -- which is assumed to correspond to the direct propagation -- is located at $t=t_0=0$. Finally, the obtained multichannel sequence $\vect{h}(t)$ is rescaled such that the vector at $t=0$ is of unit magnitude, and is hereafter referred to as ground truth RdRIR. 

The chosen evaluation metrics are aimed to reflect the algorithms' ability to recover directions and relative delays of early echoes. \revis{For that reason, we have avoided the common ``normalized project misalignement'' (NPM) error \cite{BSIEtoolbox}, which is a point-wise metric that indiscriminately penalizes even small temporal deviations from the ground truth. Instead, $\ing{N} = 15$ largest peaks of the ground truth delay-magnitude representation \eqref{eqDelMagn} are selected}, from which the corresponding delays $\{t_0, t_1, \hdots , t_{\ing{N}-1}\}$ and encoding vectors $\{\vect{h}(t_0),\vect{h}(t_1),\hdots,\vect{h}(t_{\ing{N}-1}) \}$ are logged. We independently apply the same peak-picking procedure to the delay-magnitude representation $\zeta_{\tilde{\vect{h}}}(t)$ of a given estimator $\tilde{\vect{h}}(t)$, which yields another set of delays $\{t_0, \tilde{t}_1, \hdots , \tilde{t}_{\ing{N}-1}\}$, associated with encoding vectors $\{\tilde{\vect{h}}(t_0),\tilde{\vect{h}}(\tilde{t}_1),\hdots,\tilde{\vect{h}}(\tilde{t}_{\ing{N}-1}) \}$ (note that the zero-delay vector, akin to the DoA delay $t_0$, is always retained). We keep only $\tilde{\ing{N}}$ of those estimates $\tilde{\vect{h}}_{\ing{j}} := \tilde{\vect{h}}(\tilde{t}_{\ing{j}})$ whose delays $\tilde{t}_{\ing{j}}$ are within a five-sample temporal neighborhood of the ground truth wavefronts - hence, the relative delay error tolerance is $|t_{\ing{i}} - \tilde{t}_{\ing{j}}| \leq 0.3$~ms. The percentage of retained estimates
\begin{equation*}
    \tilde{P}=\frac{\tilde{\ing{N}}}{\ing{N}}100\%,    
\end{equation*}
is to be interpreted as a detection rate indicator. Indeed, since the number of retrieved peaks is equal for both ground truth and the estimate, detection precision and recall have the same value. For all ground truth wavefronts and retained estimates, we then find the closest (in the least squared sense) SH vectors $\vect{y}_{\ing{i}} := \vect{y}(\theta_{\ing{i}},\phi_{\ing{i}})$, respectively $\tilde{\vect{y}}_{\ing{j}} =  \tilde{\vect{y}}(\tilde{\theta}_{\ing{j}},\tilde{\phi}_{\ing{j}})$, parametrized by the appropriate azimuth and elevation values. We use these directions to evaluate angular errors between an estimate and the associated ground truth wavefront:
\begin{equation*}
    \tilde{\epsilon}_{\ing{i},\ing{j}} = \angle \left( (\theta_{\ing{i}},\phi_{\ing{i}}), (\tilde{\theta}_{\ing{j}},\tilde{\phi}_{\ing{j}}) \right),
\end{equation*}
where $\angle(\cdot)$ denotes the great-circle distance for the given pair of directions. Finally, knowing that measured RIRs do not perfectly obey the structure of analytic SH vectors, we also evaluate the coherence $\tilde{c}_{\ing{i},\ing{j}}$ between a ``raw'' ground truth vector $\vect{h}_{\ing{i}} := \vect{h}(t_{\ing{i}})$, and its estimate $\tilde{\vect{h}}_{\ing{j}}$:
\begin{equation*}
    \tilde{c}_{\ing{i},\ing{j}} = \frac{\transp{\vect{h}_{\ing{i}}} \tilde{\vect{h}}_{\ing{j}} }{\norm{\vect{h}_{\ing{i}}}{} \norm{\tilde{\vect{h}}_{\ing{j}}}{} }.
\end{equation*}

\revis{
\begin{figure*}
    \centering
    \includegraphics[width=\textwidth,trim={0 1cm 0 1cm},clip]{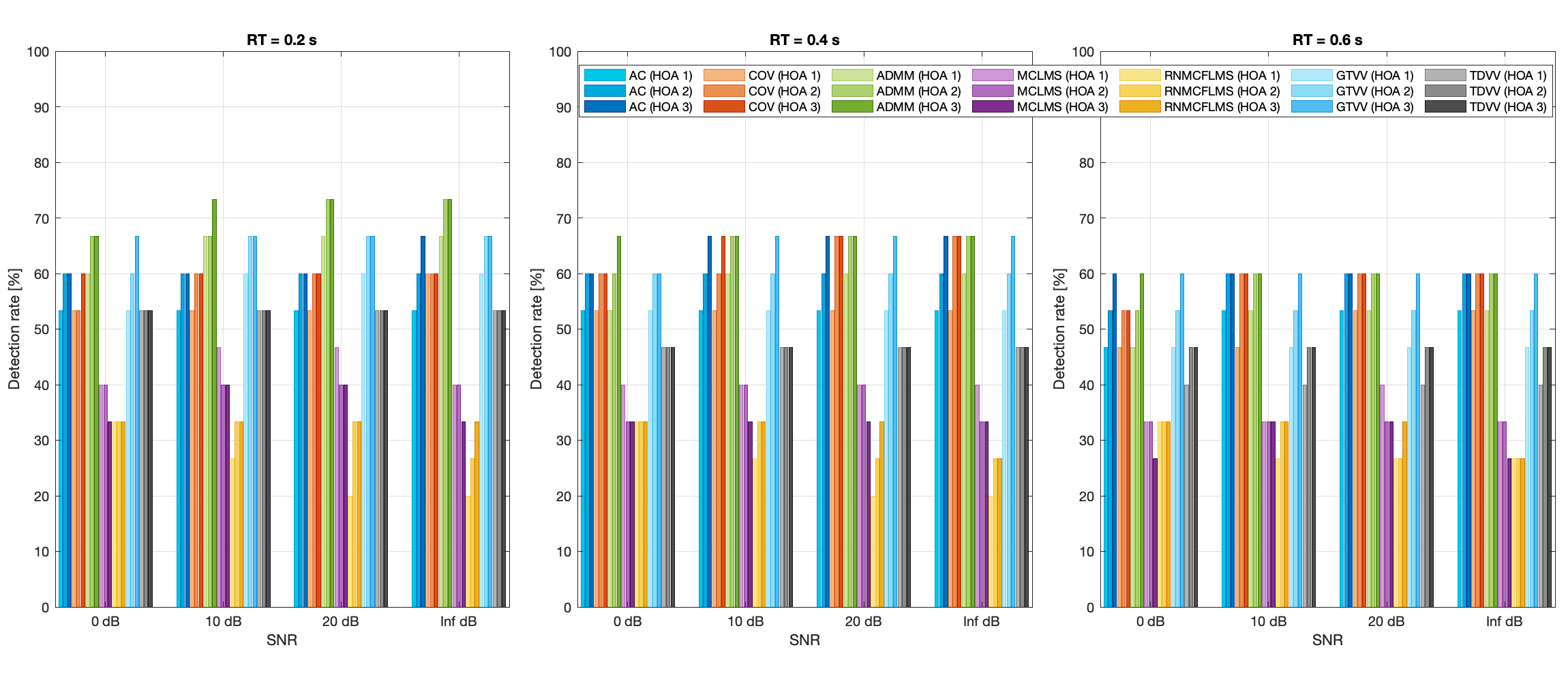} 
    \caption{\revis{Median detection rate $\tilde{P}$ results for all methods, with respect to the RT and SNR settings.}}
    \label{figDetRate}
\end{figure*}
}

\revis{Let us summarize the steps involved:
\begin{enumerate}
    \item Define the STFT frame length to be about $2\bar{\tau}_{\max}$ and compute the tensor $\hat{\vect{b}}(\ing{f},\ing{t})$.
    \item Estimate $\hat{\vect{v}}(f)$, $\vect{v}(t)$ and DoA through the ``self-steering'' procedure in Alg.~\ref{algGTVV} (for TDVV, set $\ing{num\_iter}=1$).
    \item If RdRIR estimation is used, get $\tilde{\vect{h}}(t)$ from either:
    \subitem \textbf{AC} - compute coefficients \eqref{eqACcoeffs} from $\hat{\vect{v}}^{-}_{\ing{l},:}$ and $\hat{\vect{v}}^{+}_{\ing{l},:}$, assemble and solve the linear Toeplitz system \eqref{eqLPsolution}.
    \subitem \textbf{COV} - directly compute coefficients \eqref{eqLPcoeffs}, assemble and solve the linear system \eqref{eqLPsolution}.
    \subitem \textbf{ADMM} - set the number of iterations and the parameter $\mu$, pre-compute the coefficients \eqref{eqACcoeffs_ADMM}, iterate the following steps: compute \eqref{eqProjGroupShrink} and \eqref{eqCCcoeffs_ADMM}, solve the Toeplitz system \eqref{eqOptim_a_ADMM}, compute \eqref{eqADMM3}, \eqref{eqADMM4}.
    \\
    If RdRIR is not being estimated, set $\tilde{\vect{h}}(t) =\vect{v}(t)$.
    \item Compute $\zeta_{\tilde{\vect{h}}}(t) = \norm{\tilde{\vect{h}}(t)}{2}$ and choose the delay indices of $\ing{N}$ largest peaks. Preserve only the indices within the prescribed relative delay error tolerance.
    \item For the retained indices, estimate the direction of each corresponding vector $\tilde{\vect{h}}_{\ing{j}}$:
    \subitem If RdRIR analysis has been applied, find the parameters $(\tilde{\theta}_{\ing{j}},\tilde{\phi}_{\ing{j}})$ that minimize the $\ell_2$ distance of a SH vector-valued function $\vect{y}(\theta,\phi)$ to $\tilde{\vect{h}}_{\ing{j}}$.
    \subitem If GTVV/TDVV is used directly, find the parameters $(\tilde{\theta}_{\ing{j}},\tilde{\phi}_{\ing{j}})$ that minimize \eqref{eqCost}.
    \item Calculate the performance metrics discussed above.
\end{enumerate}}

\subsection{Simulated RIRs}

In simulated experiments, we mimic the scenario where the microphone array is fixed, while the speech source is mobile, \revis{slowly moving at the average speed of $0.97$ km/h. This value is only slightly below the average speed of a moving talker in Task 3 of the LOCATA challenge \cite{evers2020locata}.} We have adapted the widely used RIR generator software \cite{habets2006room}, in order to generate Ambisonic RIRs at each of $100$ uniformly sampled positions along $10$ randomly generated smooth source trajectories. The corresponding microphone array positions are randomly chosen in the $xy$-plane, such that the array altitude is kept fixed at $1.2$~m. The microphone signals are obtained by \revis{sliding convolution and} the spatial interpolation technique implemented in the Roomsimove toolbox \cite{vincent2015roomsimove}. The virtual ``room'' has dimensions $5\times4\times3$ m\textsuperscript{3}, while the reverberation time (RT) takes values from $\{0.2\text{s},0.4\text{s},0.6\text{s}\}$. \revis{Signals are corrupted by diffuse babble noise whose impulse response has been obtained by} extracting the reverberant parts of several RIRs corresponding to random positions within the room, and then computing their average, as done in \cite{perotin2018multichannel}. The considered SNR levels are $0$, $10$, $20$ dB and noiseless (``$\infty$'' dB setting). 

\begin{figure*}
    \centering
    \includegraphics[width=\textwidth,trim={0 1cm 0 1cm},clip]{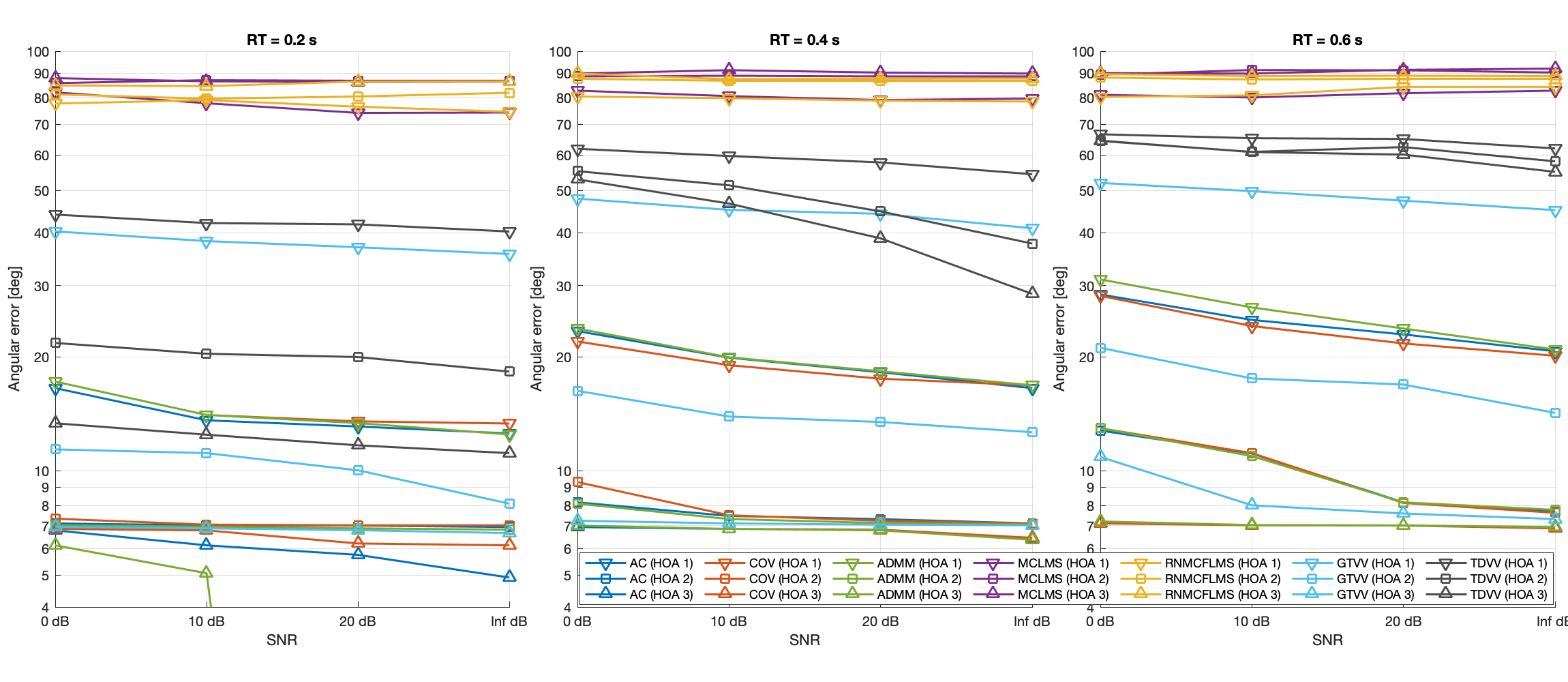}
    \caption{\revis{Median angular error for the detected wavefronts, relative to RT and SNR levels. Out-of-scope values (for ADMM (HOA 3) with SNR=20 or Inf dB, and RT=0.2s) are null.}}
    \label{figAngErr}
\end{figure*}
\begin{figure*}
    \includegraphics[width=\textwidth,trim={0 1cm 0 1cm},clip]{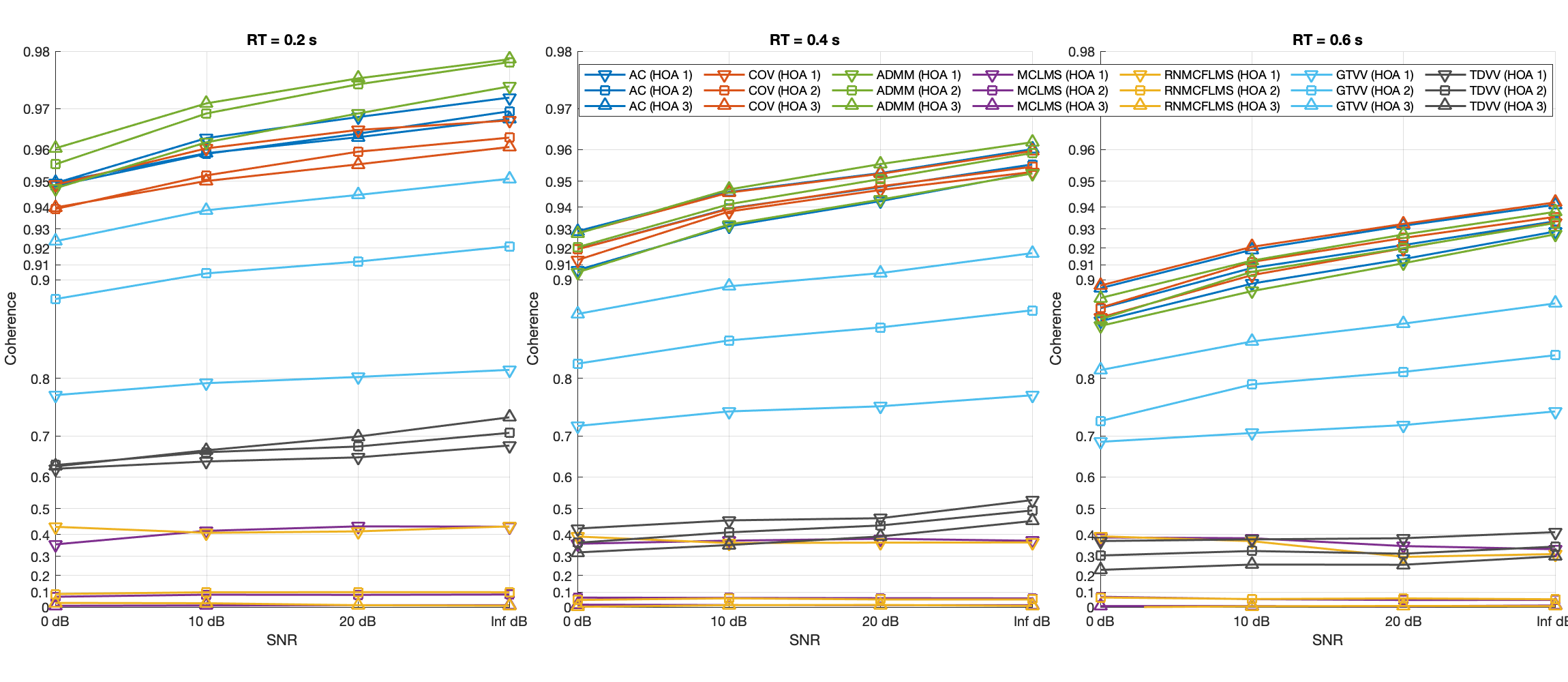}
    \caption{\revis{Median coherence for the detected wavefronts, relative RT and SNR levels.}}
    \label{figCoherence}
\end{figure*}

\revis{The reported results are computed on the ensemble of generated data, \emph{i.e.} from all generated trajectories.} For every performance metric, the experimental outputs are presented as a set of subfigures relative to each RT setting. Within a subfigure, the results are given as a function of the varying SNR level. 
The detection rate $\tilde{P}$ results are given in the form of bar plots in Fig.~\ref{figDetRate}. The RdRIR-based methods generally provide better detection than the baseline approaches, with the ADMM variant obtaining the highest percentage of accurate detections, \revis{most notably at low HOA orders}. The angular error and coherence performance on the detected wavefronts are presented in Fig.~\ref{figAngErr} and Fig.~\ref{figCoherence}, respectively. One may observe the same trend, with the proposed methods outperforming the baselines, often by a large margin. \revis{The LMS baselines performed considerably worse than the other methods, which is also reflected in their poor coherence scores in Fig.~\ref{figCoherence}. They yield essentially similar, quasi-random results across all RT and SNR levels. The obtained results were confirmed by the scores of two sample t-tests \cite{efron2012large}, evaluated for each pair of estimation methods for a given HOA order. Statistical tests also indicate that, when compared to one another, the proposed RdRIR estimation methods achieve similar performance in terms of the attained angular error and coherence. }

\revis{With the exception of the LMS baselines, the performance of all tested methods improves with the increase in SNR and HOA order}, and drops with the increase in RT. A possible remedy for the latter may be to apply the ``channel shortening'' technique, \emph{i.e.} to pre-process the input signals by a multichannel dereverberation algorithm (\emph{e.g.} \cite{yoshioka2012generalization}) before the RdRIR estimation.

\begin{figure*}
     \centering
     \begin{subfigure}[b]{0.3\textwidth}
         \centering
         \includegraphics[width=\textwidth]{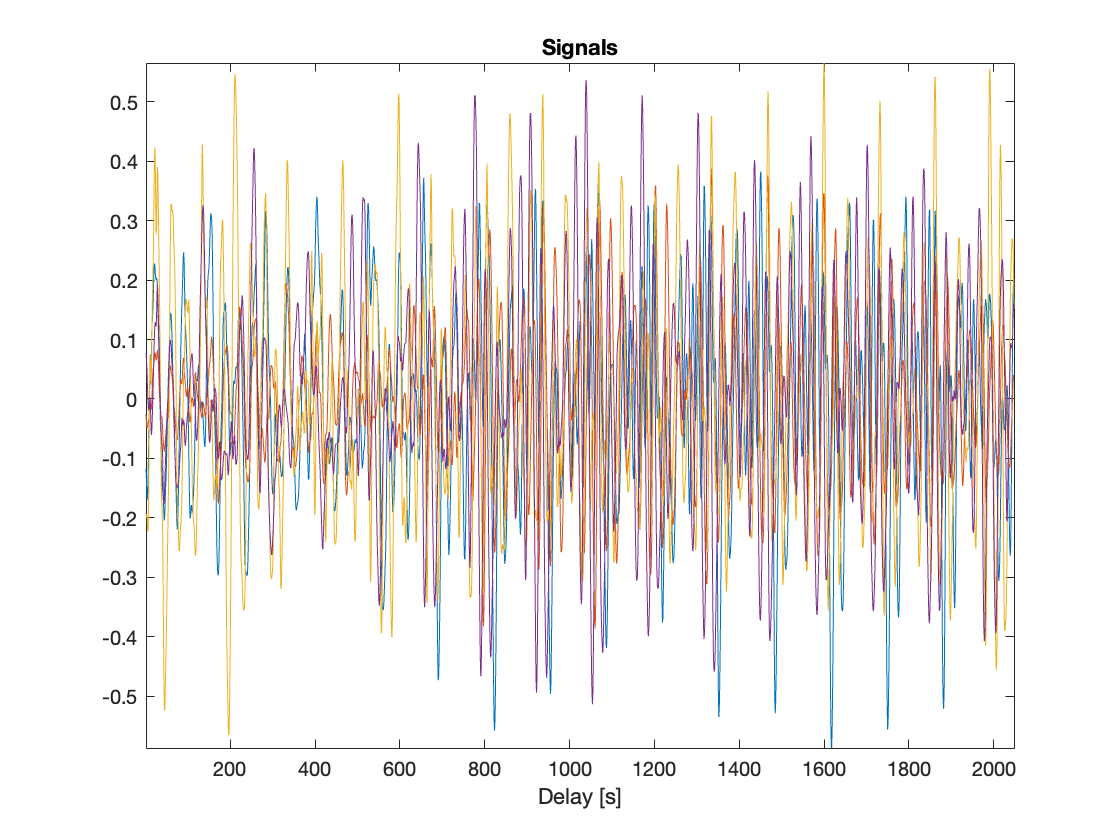}
         \caption{The (superposed) FOA channels.}\label{figExampleSignals}
     \end{subfigure}
     \begin{subfigure}[b]{0.3\textwidth}
         \centering
         \includegraphics[width=\textwidth]{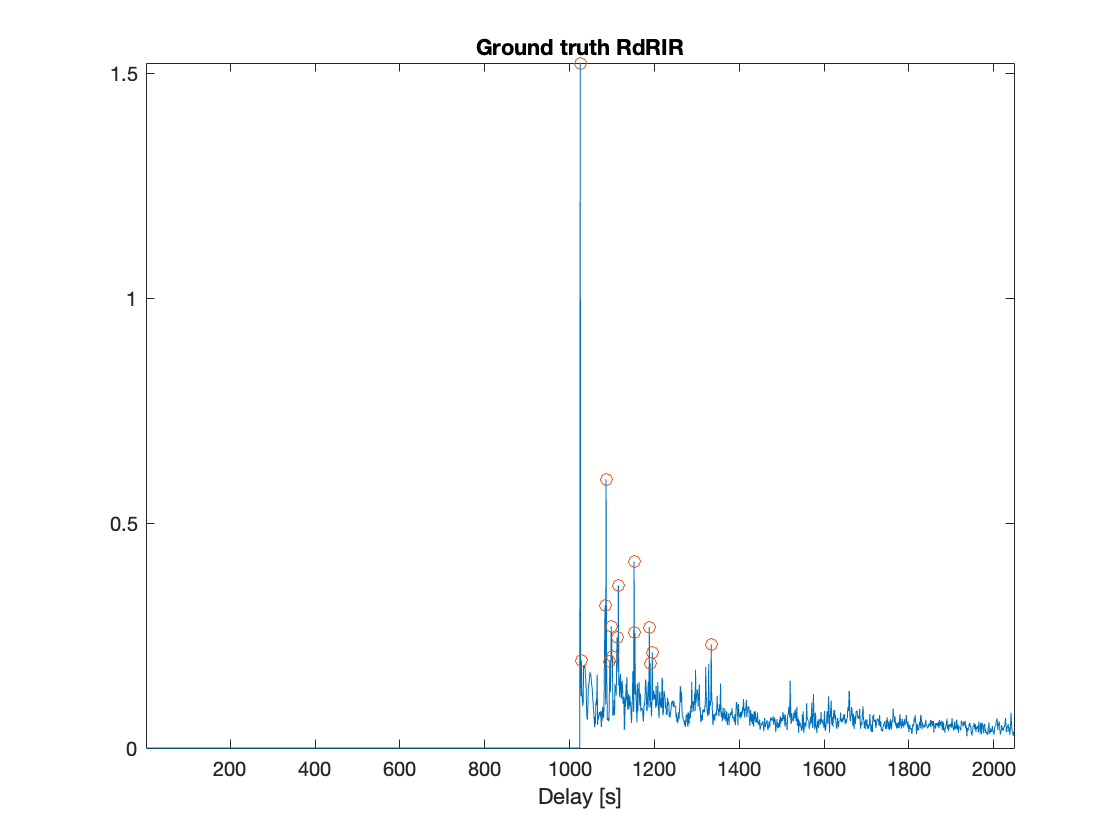}
         \caption{Ground truth RdRIR.}\label{figExampleRdRIR}
     \end{subfigure}
     \begin{subfigure}[b]{0.3\textwidth}
         \centering
         \includegraphics[width=\textwidth]{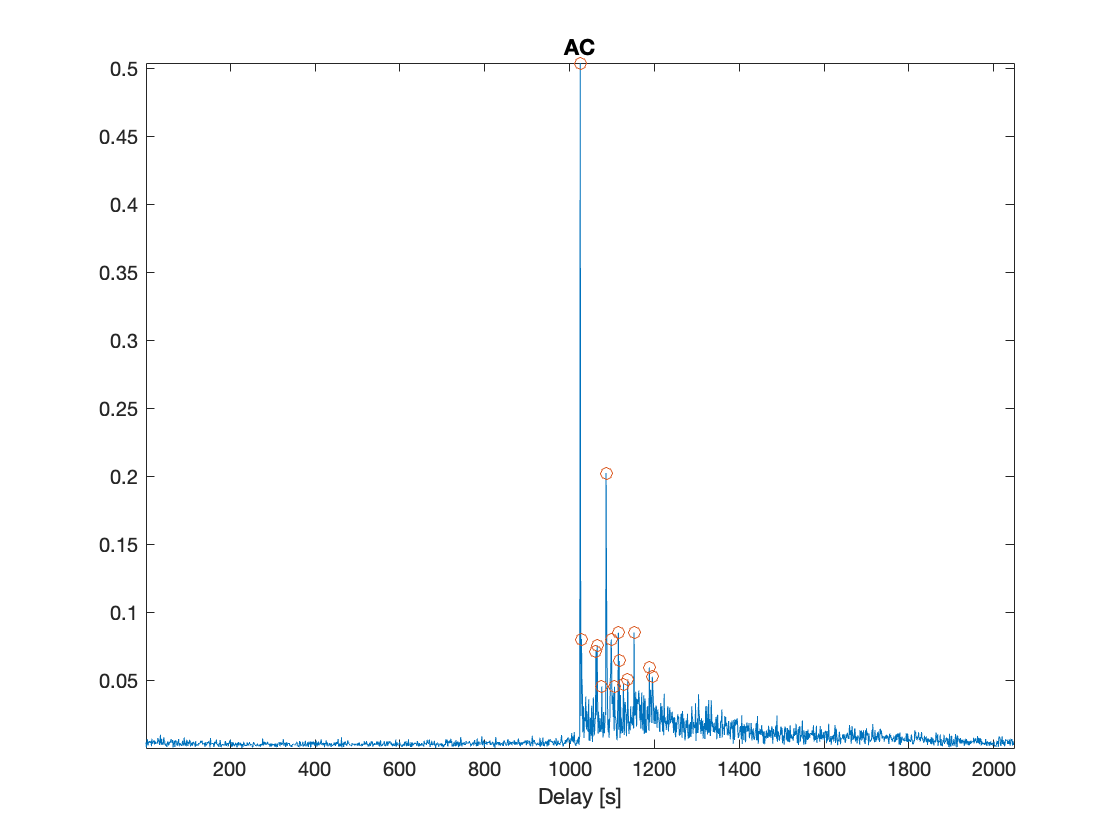}
         \caption{AC method}\label{figExampleAC}
     \end{subfigure}
     \begin{subfigure}[b]{0.3\textwidth}
         \centering
         \includegraphics[width=\textwidth]{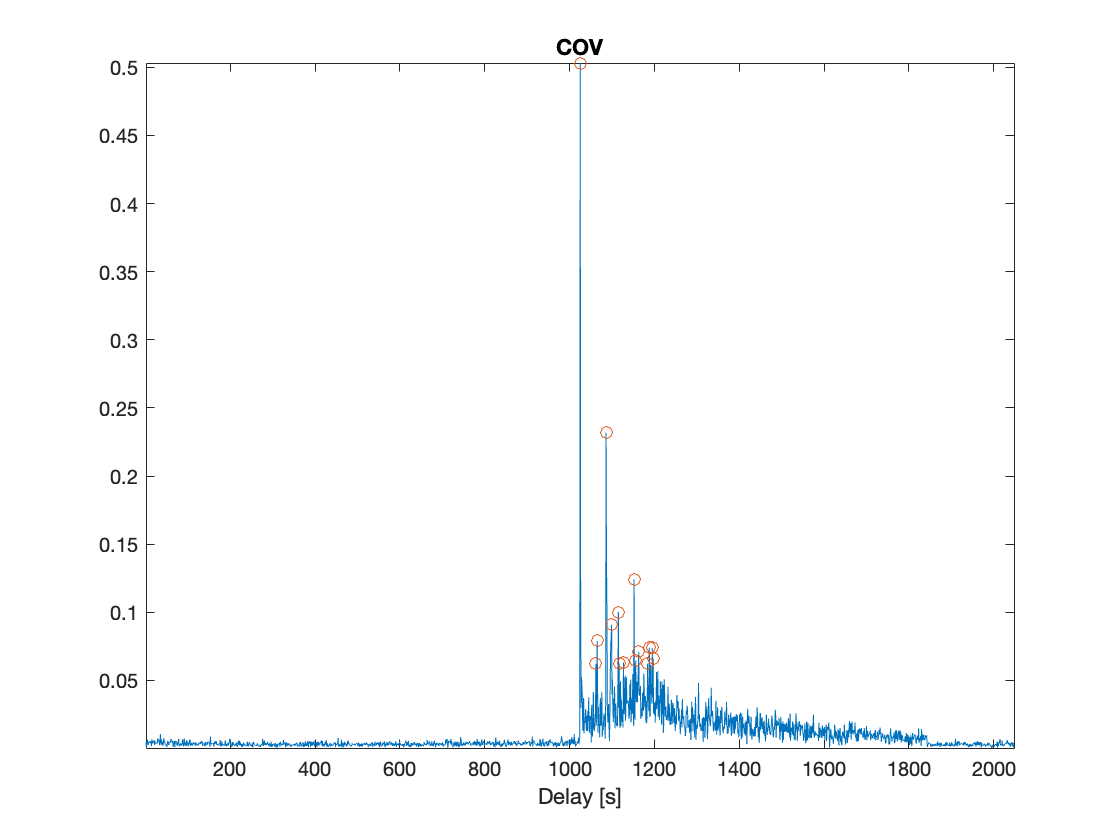}
         \caption{COV method.}\label{figExampleCOV}
     \end{subfigure}
     \begin{subfigure}[b]{0.3\textwidth}
         \centering
         \includegraphics[width=\textwidth]{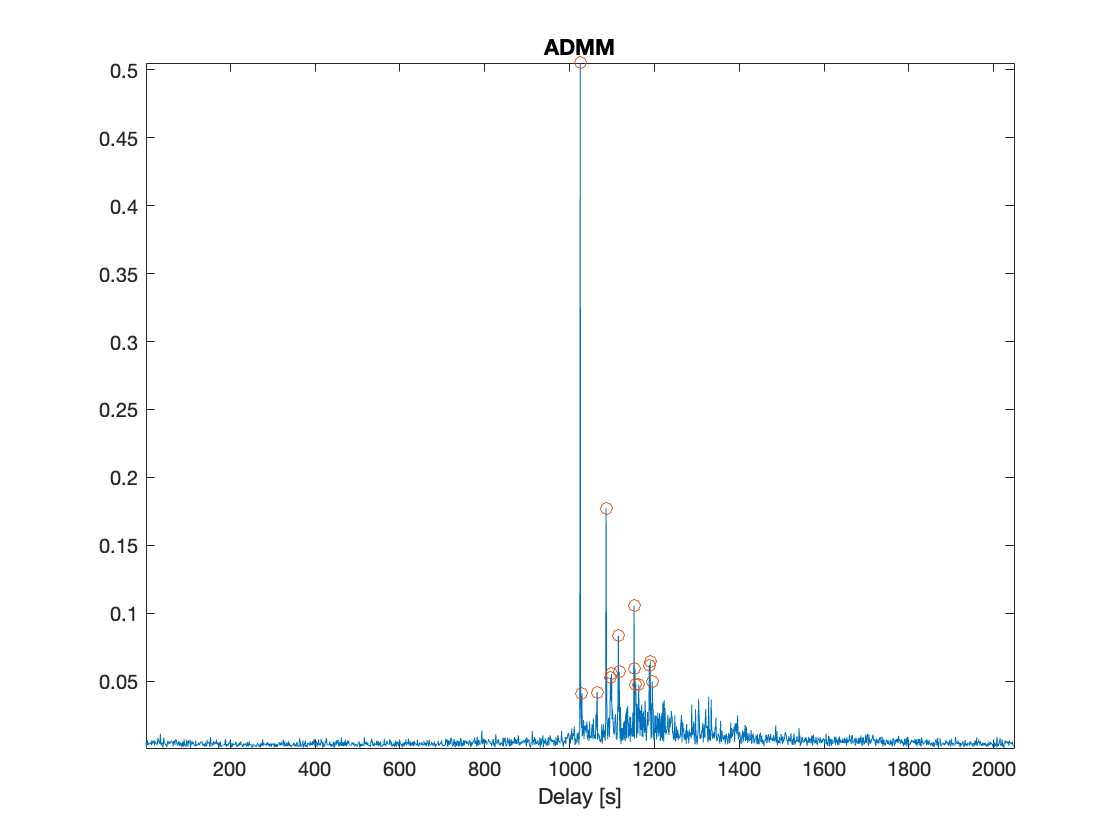}
         \caption{ADMM method.}\label{figExampleADMM}
     \end{subfigure}
     \begin{subfigure}[b]{0.3\textwidth}
         \centering
         \includegraphics[width=\textwidth]{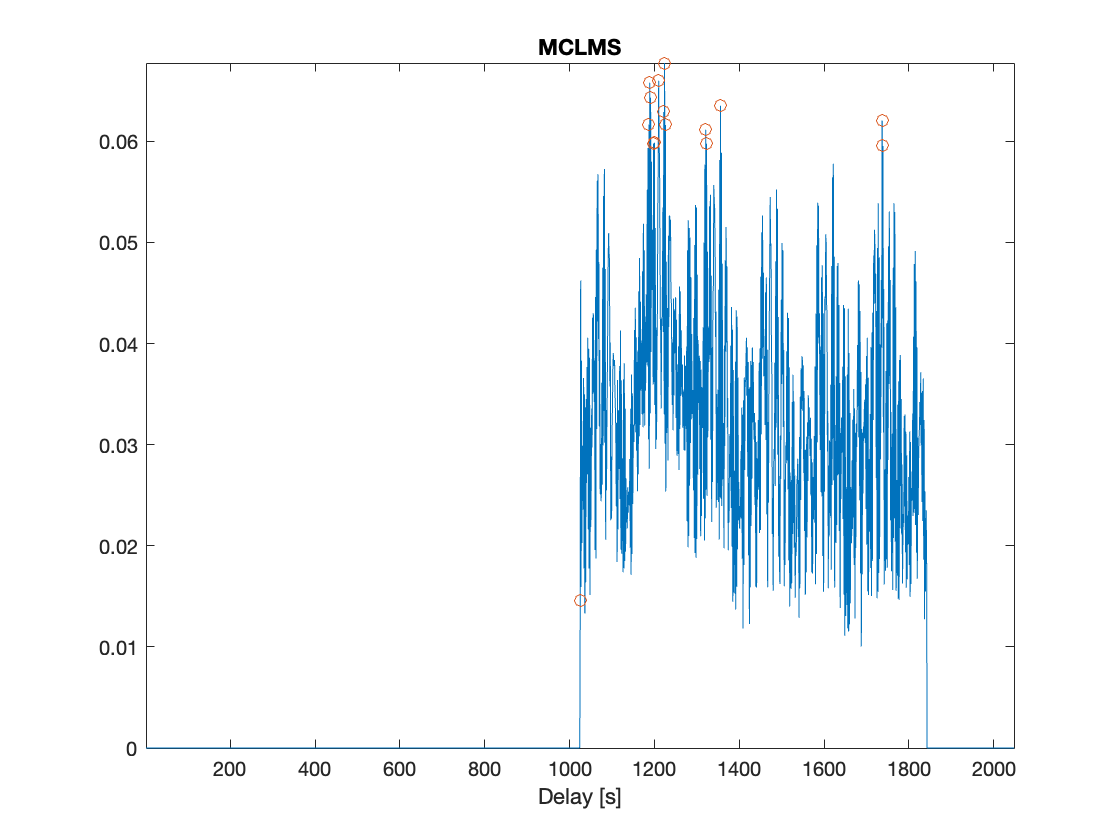}
         \caption{MCLMS method.}\label{figExampleMCLMS}
     \end{subfigure}
     \begin{subfigure}[b]{0.3\textwidth}
         \centering
         \includegraphics[width=\textwidth]{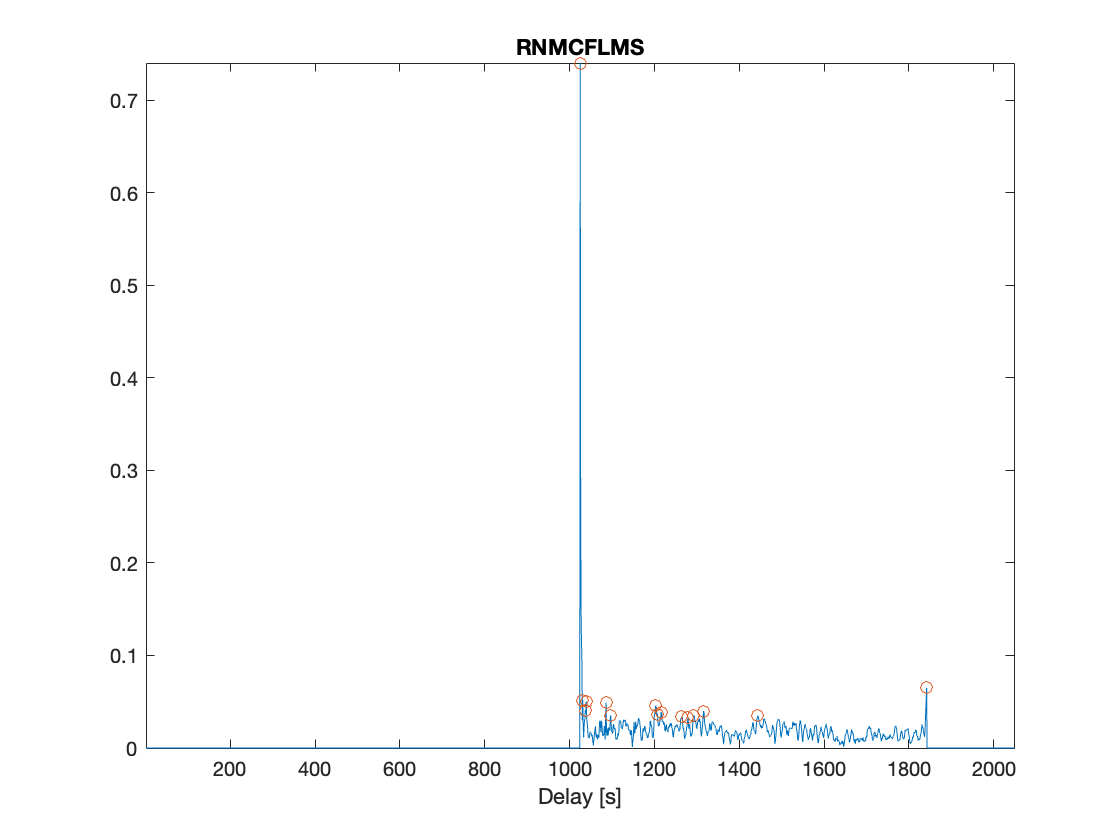}
         \caption{RNMCFLMS method.}\label{figExampleRNMCFLMS}
     \end{subfigure}
     \begin{subfigure}[b]{0.3\textwidth}
         \centering
         \includegraphics[width=\textwidth]{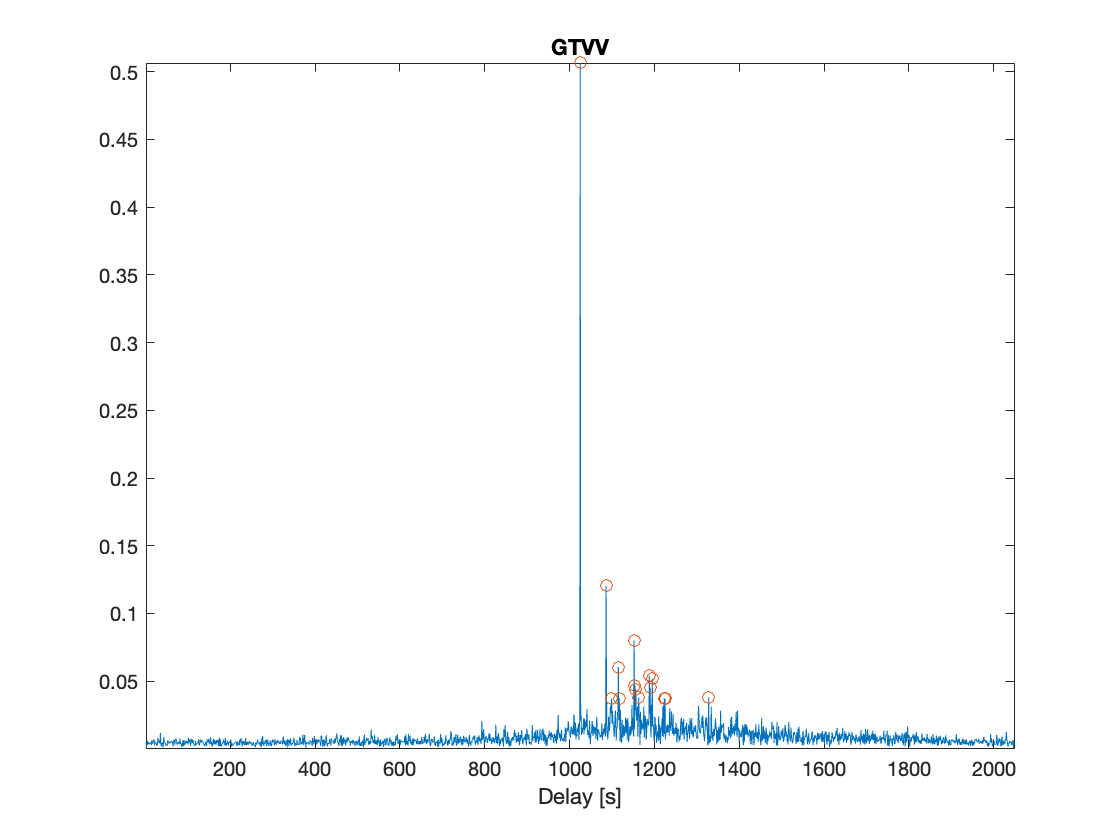}
         \caption{GTVV representation.}\label{figExampleGTVV}
     \end{subfigure}
     \begin{subfigure}[b]{0.3\textwidth}
         \centering
         \includegraphics[width=\textwidth]{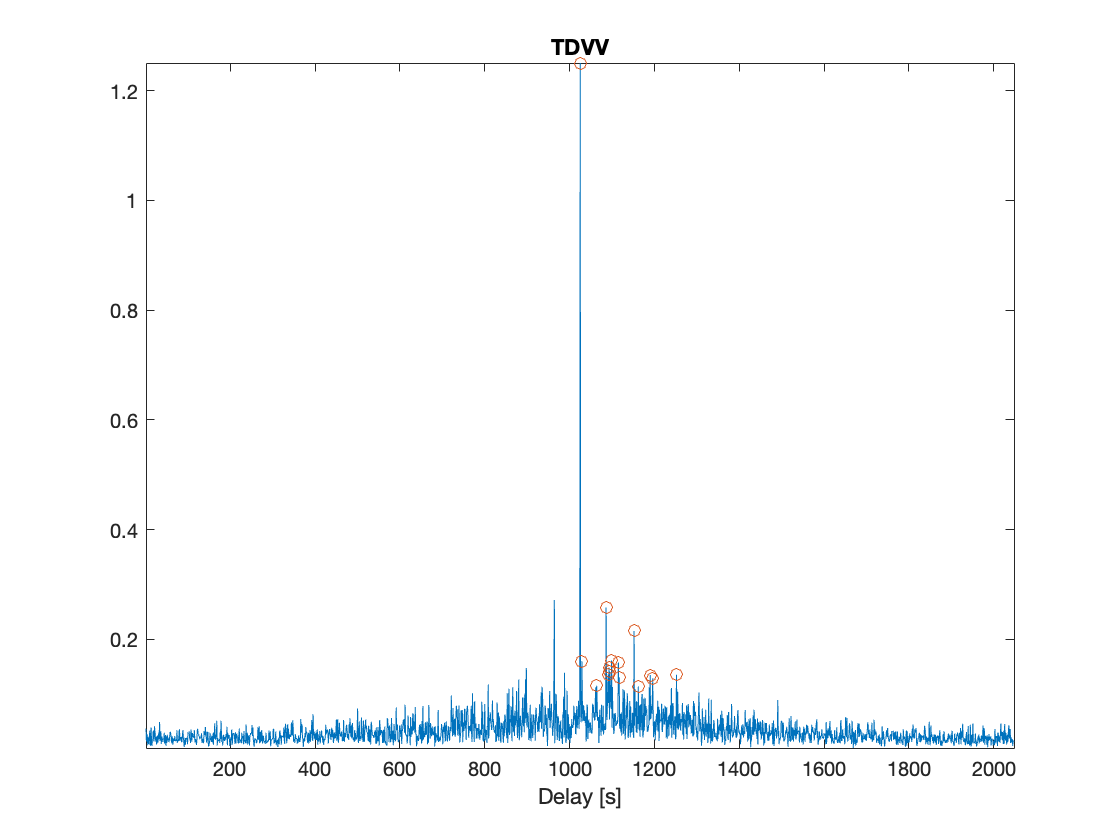}
         \caption{TDVV representation.}\label{figExampleTDVV}
     \end{subfigure} 
     \caption{Delay-magnitude representations of the estimates for all tested methods (Figs.~\ref{figExampleAC}-\ref{figExampleTDVV}), from the FOA recording in Fig.~\ref{figExampleSignals}. The ground truth RdRIR is given in Fig.~\ref{figExampleRdRIR}, and the selected peaks are denoted by red circles.}\label{figExampleAll}
\end{figure*}

\subsection{Recorded RIRs}

In order to evaluate the performance of proposed methods in more realistic conditions, we use the dataset of recorded Ambisonic RIRs from University of Aalto, Finland \cite{mckenzie2021dataset}. The dataset contains RIRs in different reverberation conditions, obtained by varying acoustic absorbers in $5$ steps, 
from mild reverberation  \revis{($T_{20}$ around $0.37$ s) to highly reverberant ($T_{20}$ about $1.21$ s)}. The authors have recorded RIRs corresponding to all combinations of $3$ positions of a sound source (Genelec 8331A coaxial loudspeaker), and $7$ positions of a microphone array (mhAcoustics Eigenmike\textregistered\;em32 and Zylia ZM-1). However, we do not consider the two microphone positions for which the source is facing the opposite direction. Instead, we use these to generate the diffuse impulse response for the additive babble noise, as discussed in the previous subsection. \revis{In this series of experiments, the SNR is fixed to $20$ dB (investigating the robustness of the proposed methods under different SNRs, on real data, is left for future work)}.

\revis{For visual comparison, examples of the obtained delay-magnitude representations, along with the corresponding input FOA signals, are presented in Fig.~\ref{figExampleAll} (note that each of the four FOA channels in the subfigure~\ref{figExampleSignals} is given in a different color). While for the proposed AC, COV and ADMM methods (and even the GTVV representation), these strongly resemble the ground truth, for TDVV and the LMS baselines this is clearly not the case.}


\begin{table*}[!ht]
    \caption{Angular error / coherence / detection rate of the signals obtained using RIRs recoded by Zylia SMA.}\label{tabZylia}
    \centering
    \begin{tabular}{|c|l|l|l|l|l|l|}
    \hline
        HOA & Method & $T_{20} = 1.21$ s & $T_{20} = 0.77$ s & $T_{20} = 0.57$ s & $T_{20} = 0.45$ s & $T_{20} = 0.37 $s \\ \hline \hline
        \multirow{7}{*}{1} & AC & \textbf{44.75°} / \textbf{0.71} / \textbf{47\%} & 40.27° / 0.81 / \textbf{47\%} & 39.53° / \textbf{0.83} / 47\% & 35.32° / 0.83 / 47\% & 32.35° / 0.82 / 47\% \\
        \cline{2-7} & COV & 44.82° / \textbf{0.71} / \textbf{47\%} & \textbf{38.72°} / \textbf{0.83} / \textbf{47\%} & \textbf{38.04°} / 0.81 / 47\% & \textbf{33.79°} / \textbf{0.83} / \textbf{53\%} & \textbf{30.80°} / \textbf{0.85} / 47\% \\ 
        \cline{2-7} & ADMM & 46.69° / 0.70 / \textbf{47\%} & 40.75° / 0.81 / \textbf{47\%} & 44.14° / 0.78 / \textbf{53\%} & 40.46° / 0.78 / \textbf{53\%} & 38.28° / 0.80 / \textbf{53\%} \\ 
        \cline{2-7} & MCLMS & 84.82° / 0.31 / 7\% & 85.95° / 0.30 / 7\% & 84.66° / 0.32 / 7\% & 80.4° / 0.35 / 7\% & 84.32° / 0.30 / 7\% \\ 
        \cline{2-7} & RNMCFLMS & 85.00° / 0.26 / 27\% & 83.08° / 0.28 / 27\% & 81.57° / 0.38 / 27\% & 82.19° / 0.28 / 20\% & 88.92° / 0.26 / 20\% \\ 
        \cline{2-7} & GTVV & 59.11° / 0.55 / 40\% & 61.43° / 0.63 / 40\% & 65.50° / 0.57 / 47\% & 61.37° / 0.63 / 47\% & 60.10° / 0.47 / 47\% \\ 
        \cline{2-7} & TDVV & 73.4° / 0.30 / 33\% & 67.85° / 0.44 / 33\% & 74.29° / 0.29 / 40\% & 70.89° / 0.23 / 40\% & 63.86° / 0.35 / 40\% \\ \hline \hline

        \multirow{7}{*}{2} & AC & \textbf{33.72°} / \textbf{0.58} / 47\% & \textbf{21.30°} / \textbf{0.77} / 47\% & \textbf{25.11°} / \textbf{0.72} / \textbf{53\%} & 26.33° / 0.73 / \textbf{53\%} & 17.63° / 0.76 / \textbf{53\%} \\
        \cline{2-7} & COV & 37.48° / 0.57 / \textbf{53\%} & 21.82° / 0.75 / \textbf{53\%} & 26.17° / 0.69 / \textbf{53\%} & 25.55° / 0.69 / \textbf{53\%} & \textbf{17.62°} / 0.78 / \textbf{53\%} \\
        \cline{2-7} & ADMM & \textbf{33.72°} / 0.57 / 47\% & 22.9° / \textbf{0.77} / \textbf{53\%} & 28.36° / \textbf{0.72} / \textbf{53\%} & \textbf{24.68°} / \textbf{0.76} / \textbf{53\%} & 18.69° / \textbf{0.79} / \textbf{53\%} \\
        \cline{2-7} & MCLMS & 83.23° / 0.15 / 7\% & 84.32° / 0.15 / 7\% & 77.11° / 0.20 / 7\% & 73.67° / 0.21 / 7\% & 74.01° / 0.18 / 7\% \\
        \cline{2-7} & RNMCFLMS & 101.93° / -0.07 / 27\% & 101.93° / -0.02 / 27\% & 100.7° / -0.01 / 33\% & 98.78° / 0.10 / 33\% & 102.83° / 0.04 / 33\% \\
        \cline{2-7} & GTVV & 56.07° / 0.41 / 47\% & 40.14° / 0.56 / 40\% & 47.87° / 0.54 / 47\% & 44.69° / 0.62 / \textbf{53\%} & 40.27° / 0.54 / 47\% \\
        \cline{2-7} & TDVV & 67.13° / 0.25 / 40\% & 62.60° / 0.36 / 40\% & 72.56° / 0.17 / 40\% & 73.20° / 0.18 / 47\% & 59.03° / 0.27 / 47\% \\ \hline \hline

        \multirow{7}{*}{3} & AC & 17.32° / 0.65 / \textbf{53\%} & \textbf{11.02°} / \textbf{0.80} / \textbf{53\%} & \textbf{13.56°} / \textbf{0.77} / \textbf{60\%} & 13.46° / 0.77 / \textbf{60\%} & 11.58° / 0.79 / \textbf{60\%} \\
        \cline{2-7} & COV & 16.92° / \textbf{0.66} / \textbf{53\%} & 11.29° / 0.77 / \textbf{53\%} & 14.34° / 0.74 / 53\% & 14.77° / 0.74 / \textbf{60\%} & 11.34° / 0.78 / \textbf{60\%} \\
        \cline{2-7} & ADMM & \textbf{14.25°} / \textbf{0.66} / \textbf{53\%} & 11.40° / \textbf{0.80} / \textbf{53\%} & 14.50° / \textbf{0.77} / \textbf{60\%} & \textbf{12.53°} / \textbf{0.82} / \textbf{60\%} & \textbf{8.16°} / \textbf{0.83} / \textbf{60\%} \\
        \cline{2-7} & MCLMS & 80.22° / 0.13 / 7\% & 83.16° / 0.10 / 7\% & 76.9° / 0.13 / 7\% & 70.84° / 0.12 / 7\% & 69.32° / 0.13 / 7\% \\
        \cline{2-7} & RNMCFLMS & 98.86° / 0.04 / 27\% & 98.86° / 0.03 / 27\% & 98.86° / 0.11 / 33\% & 99.97° / 0.10 / 33\% & 101.21° / 0.06 / 33\% \\
        \cline{2-7} & GTVV & 39.49° / 0.47 / 47\% & 21.62° / 0.61 / 47\% & 31.75° / 0.59 / 47\% & 18.47° / 0.69 / 53\% & 20.82° / 0.62 / 53\% \\
        \cline{2-7} & TDVV & 68.41° / 0.23 / 40\% & 64.22° / 0.31 / 40\% & 66.92° / 0.20 / 47\% & 66.77° / 0.19 / 47\% & 56.69° / 0.24 / 47\% \\ \hline

    \end{tabular}
\end{table*}

\begin{table*}[!ht]
    \centering
    \caption{Angular error / coherence / detection rate of the signals obtained using RIRs recoded by Eigenmike SMA.}\label{tabEigenmike}
    \begin{tabular}{|c|l|l|l|l|l|l|}
    \hline
        HOA & Method & $T_{20} = 1.21$ s & $T_{20} = 0.77$ s & $T_{20} = 0.57$ s & $T_{20} = 0.45$ s & $T_{20} = 0.37 $s \\ \hline \hline
        \multirow{7}{*}{1} & AC & 19.71° / \textbf{0.91} / \textbf{53\%} & 19.71° / \textbf{0.90} / \textbf{47\%} & \textbf{19.88°} / \textbf{0.90} / \textbf{53\%} & 18.38° / \textbf{0.92} / \textbf{53\%} & 19.99° / \textbf{0.92} / \textbf{53\%} \\
        \cline{2-7} & COV & \textbf{19.57°} / \textbf{0.91} / 47\% & \textbf{19.29°} / \textbf{0.90} / \textbf{47\%} & 20.24° / \textbf{0.90} / 47\% & \textbf{18.35°} / \textbf{0.92} / 47\% & \textbf{19.85°} / 0.91 / \textbf{53\%} \\ 
        \cline{2-7} & ADMM & 19.58° / 0.89 / \textbf{53\%} & 21.06° / 0.87 / \textbf{47\%} & 20.64° / \textbf{0.90} / 47\% & 20.63° / 0.89 / 47\% & 19.91° / 0.90 / \textbf{53\%} \\ 
        \cline{2-7} & MCLMS & 84.97° / 0.30 / 7\% & 82.66° / 0.32 / 7\% & 79.99° / 0.36 / 7\% & 81.26° / 0.31 / 7\% & 78.87° / 0.36 / 7\% \\ 
        \cline{2-7} & RNMCFLMS & 73.75° / 0.40 / 33\% & 69.41° / 0.42 / 33\% & 61.76° / 0.58 / 27\% & 68.05° / 0.50 / 27\% & 57.21° / 0.60 / 27\% \\ 
        \cline{2-7} & GTVV & 39.09° / 0.78 / 47\% & 45.46° / 0.69 / 40\% & 40.43° / 0.74 / 47\% & 37.42° / 0.78 / 53\% & 36.34° / 0.80 / 47\% \\ 
        \cline{2-7} & TDVV & 46.29° / 0.51 / 47\% & 48.57° / 0.49 / 40\% & 50.68° / 0.46 / 47\% & 49.92° / 0.49 / 40\% & 46.84° / 0.52 / 47\% \\ \hline \hline

        \multirow{7}{*}{2} & AC & 7.39° / 0.90 / \textbf{60\%} & \textbf{7.77°} / \textbf{0.88} / \textbf{53\%} & 8.01° / \textbf{0.89} / \textbf{60\%} & 8.21° / \textbf{0.90} / \textbf{60\%} & 11.39° / \textbf{0.89} / \textbf{60\%} \\ 
        \cline{2-7} & COV & 7.39° / 0.90 / 53\% & 7.91° / \textbf{0.88} / \textbf{53\%} & 8.01° / 0.88 / 53\% & 8.21° / \textbf{0.90} / \textbf{60\%} & 11.7° / \textbf{0.89} / 53\% \\ 
        \cline{2-7} & ADMM & \textbf{7.23°} / \textbf{0.91} / 53\% & \textbf{7.77°} / 0.86 / \textbf{53\%} & \textbf{7.53°} / \textbf{0.89} / 53\% & \textbf{7.67°} / 0.89 / 53\% & \textbf{8.16°} / 0.88 / 53\% \\ 
        \cline{2-7} & MCLMS & 80.35° / 0.09 / 7\% & 84.04° / 0.10 / 7\% & 70.14° / 0.17 / 7\% & 79.93° / 0.16 / 7\% & 64.91° / 0.26 / 7\% \\ 
        \cline{2-7} & RNMCFLMS & 80.54° / 0.13 / 40\% & 79.50° / 0.09 / 40\% & 75.43° / 0.17 / 40\% & 73.37° / 0.20 / 40\% & 71.56° / 0.24 / 40\% \\ 
        \cline{2-7} & GTVV & 13.02° / 0.84 / 53\% & 15.29° / 0.79 / \textbf{53\%} & 13.25° / 0.83 / 53\% & 13.28° / 0.83 / 53\% & 13.95° / 0.84 / 53\% \\ 
        \cline{2-7} & TDVV & 35.36° / 0.44 / 47\% & 39.87° / 0.38 / 47\% & 33.85° / 0.43 / 53\% & 38.62° / 0.45 / 47\% & 38.68° / 0.47 / 53\% \\ \hline \hline

        \multirow{7}{*}{3} & AC & 7.05° / 0.89 / 53\% & \textbf{7.11°} / \textbf{0.87} / \textbf{53\%} & \textbf{7.05°} / \textbf{0.87} / \textbf{60\%} & \textbf{7.12°} / \textbf{0.87} / \textbf{60\%} & 7.31° / 0.85 / \textbf{60\%} \\ 
        \cline{2-7} & COV & 7.11° / 0.89 / 53\% & 7.19° / 0.86 / \textbf{53\%} & 7.12° / 0.85 / 53\% & 7.20° / 0.86 / \textbf{60\%} & 7.31° / 0.85 / \textbf{60\%} \\ 
        \cline{2-7} & ADMM & \textbf{7.02°} / \textbf{0.90} / 53\% & 7.19° / 0.86 / \textbf{53\%} & 7.20° / \textbf{0.87} / 53\% & 7.23° / \textbf{0.87} / \textbf{60\%} & \textbf{7.20°} / \textbf{0.87} / \textbf{60\%} \\ 
        \cline{2-7} & MCLMS & 82.10° / 0.07 / 7\% & 78.31° / 0.09 / 7\% & 72.41° / 0.12 / 7\% & 73.13° / 0.11 / 7\% & 66.27° / 0.17 / 7\% \\ 
        \cline{2-7} & RNMCFLMS & 86.34° / 0.12 / 33\% & 84.39° / 0.10 / 33\% & 80.95° / 0.09 / 40\% & 82.09° / 0.13 / 40\% & 79.84° / 0.15 / 40\% \\ 
        \cline{2-7} & GTVV & 7.47° / 0.85 / \textbf{60\%} & 7.77° / 0.81 / \textbf{53\%} & 7.77° / 0.82 / \textbf{60\%} & 8.21° / 0.80 / 53\% & 7.62° / 0.83 / 53\% \\ 
        \cline{2-7} & TDVV & 30.6° / 0.36 / 47\% & 39.62° / 0.31 / 47\% & 30.17° / 0.31 / 53\% & 31.18° / 0.35 / 53\% & 32.95° / 0.38 / 53\% \\ \hline

    \end{tabular}
\end{table*}

The results are given in Tables \ref{tabEigenmike} and \ref{tabZylia}, as a function of HOA order, $T_{20}$ and the type of spherical microphone array (SMA) used for recording RIRs. The best results are emphasized by the boldface font. The proposed methods clearly outperform the baselines, in terms of all evaluation metrics. The three RdRIR-based approaches obtain comparable results, without a clear winner in terms of estimation performance. However, given that the AC method is the least computationally demanding, it seems to be best suited for practical applications. As expected, amongst baseline approaches, the GTVV representation produces the best results, especially for higher HOA orders. One may also remark that the overall performance of all tested methods improves with the HOA order, and -- somewhat surprisingly -- is not much affected by the change in sound absorption, \emph{i.e.}, by the RT of the room. 

It is important to indicate that some degradation in performance (particularly, in detection rate) may be due to the chosen experimentation protocol, based on pre-selected peaks of the ground truth RdRIR. Related to that, note that while the angular precision and coherence are generally correlated, this is not always the case, suggesting that a more refined method for extracting directions from SH vectors (such as \cite{herzog2019eigenbeam}) may further reduce angular errors. This may also explain, to some extent, the disparity between the results obtained from the Eigenmike and Zylia SMAs.

\section{CONCLUSION}
\label{secConclusion}

We have presented a detailed discussion on GTVV -- Generalized Time-domain Velocity Vector -- and proposed several methods for the blind identification of early room impulse responses by exploiting properties of this signal representation. We term the time series extracted by these methods \emph{RdRIR - Reduced Room Impulse Response}. The numerical experiments using simulated and recorded RIRs (acquired by different SMAs) demonstrate the performance gains of RdRIR over the baseline BSI approaches. We envision that some of the proposed techniques, due to their implementation simplicity and small computational overhead, could find their place in many practical applications involving Ambisonics and immersive sound. Future work will focus on improving the angular precision of estimated wavefronts, use of RdRIR representation in learned models (\emph{e.g.} deep neural networks), \revis{support for multiple sound sources}, and potentially, on extending the benefits of RdRIR beyond Ambisonics, or even the spatial audio context itself.



{\appendices
\section{Relation with pseudointensity vector}\label{appPIV}

Sound intensity is defined as the product of acoustic pressure and particle velocity \cite{jacobsen1991note,kuttruff2016room}. In a pure-sound field \cite{jacobsen1991note}, its real part -- \emph{active} sound intensity -- is orthogonal to the incoming wavefront, suggesting it can be used for determining DoA. The linearized fluid momentum equation states that particle velocity is aligned with the spatial gradient of acoustic pressure \cite{merimaa2006analysis}, hence one needs only an estimate of the acoustic pressure and its gradient to approximate this quantity. Pseudointensity vector is an FOA approximation of active sound intensity, defined as \cite{jarrett20103d,merimaa2006analysis}
\begin{equation}
  \hat{\vect{i}}(f) = \Re\left(\hat{b}_0(f)^* \hat{\vect{b}}_{1:3}(f) \right),
\end{equation}
where $\Re$ denotes the real part of a complex number, $\hat{b}_0(f)$ is the first (omnidirectional) channel, while $\hat{\vect{b}}_{1:3}(f)$ is the vector of the remaining three FOA channels. Indeed, while $\hat{b}_0(f)$ is a good approximation of acoustic pressure at the center of an array \cite{rafaely2015fundamentals}, the other FOA channels exhibit spatial response similar to figure-of-eight microphones aligned with Cartesian coordinate axes \cite{zotter2019ambisonics}. Hence, $\hat{\vect{b}}_{1:3}$ is a decent approximation of the spatial gradient vector. 

Note that, for the trivial beamformer $\vect{w} = \transp{\left[ \begin{smallmatrix} 1 & 0 & 0 & \hdots & 0 \end{smallmatrix} \right]}$, we can express $\hat{\vect{i}}(f)$ using GFVV \eqref{eqFDVVinst}, as follows:
\begin{equation}\label{eqPseudointensity}
  \hat{\vect{i}}(f) = \frac{1}{|\hat{b}_0(f)|^2}\Re\left( \left[ \begin{matrix} \hat{v}_1(f) \\ \hat{v}_2(f) \\ \hat{v}_3(f) \end{matrix} \right] \right).
\end{equation}
Thus, $\hat{\vect{i}}(f)$ is parallel to the real part of the RTF vector, excluding its first entry (which is trivially equal to $1$). 

Consider a very simple scenario: in addition to the wavefront coming from the DoA direction $(\theta_0,\phi_0)$, there is an impinging wavefront from a reflected sound in the direction $(\theta_1,\phi_1)$. Given the unit response of the omnidirectional channel in all directions, from \eqref{eqFDVVinst} we can rewrite \eqref{eqPseudointensity} as 
\begin{equation}
  \hat{\vect{i}}(f) \propto \Re \left( \frac{\vec{u}_0 - \vec{u}_1 \gamma_1}{1 - \gamma_1} \right), 
\end{equation}
where $\gamma_1=-\hat{g}_1(f)e^{-j2\pi f \tau_1}$, while we use $\vec{u}_0$ and $\vec{u}_1$ to designate the subvectors composed of entries $\transp{\left[y[1],y[2],y[3]\right]}$ of the SH encoding vectors $\vect{y}_0$ and $\vect{y}_1$, respectively.

Disregarding the scaling factors, we have 
\begin{equation}
  \hat{\vect{i}}(f) \propto \vec{u}_0 \left(1 + \hat{g}_1(f) \cos \varphi_1 \right) + \vec{u}_1 \hat{g}_1(f)\left( \hat{g}_1(f) + \cos \varphi_1 \right),
\end{equation}
where $\varphi_1 = 2 \pi f \tau_1$. Since $\varphi_1$ varies linearly along frequencies $f$, we generally have $\cos (2 \pi f \tau_1) \neq -\hat{g}_1(f)$, hence the pseudointensity vector $\hat{\vect{i}}(f)$ produces a biased estimate of the DoA direction $\vec{u}_0$. At the same time, due to \eqref{eqTaylorConditionStronger} and the assumed $\hat{g}_1(f)<1$, the DoA estimate from GTVV (or even TDVV) representation, would remain unbiased.

\section{ADMM for the problem \eqref{EQOPTIM_AH}}\label{appADMM}

With $\mu > 0$, $\iter{\tilde{\mtrx{H}}}{0} = \iter{\mtrx{U}}{0} = \mtrx{0}$, and $\mtrx{0} \Rset{\ing{(L+1)}^2}{\ing{J}}$ the all-zero matrix, the proposed ADMM iterates the next steps:
\begin{align}
  \iter{\mtrx{H}}{q+1} & = \argmin_{\mtrx{H} \in \Xi} \mu\revis{\norm{\mtrx{H}}{2,1}} + \frac{1}{2}\norm{\mtrx{H} - \iter{\tilde{\mtrx{H}}}{q} - \iter{\mtrx{U}}{q} }{F}^2 \label{eqADMM1} \\
  \iter{\vect{a}}{q+1} & = \argmin_{\vect{a}, \; a_0=1} \sum\limits_{\ing{l}=0}^{(\ing{L}+1)^2-1} \norm{\vect{v}_{\ing{l},:} \ast \vect{a} - \iter{\vect{h}}{q+1}_{\ing{l},:} + \iter{\vect{u}}{q}_{\ing{l},:} }{2}^2 \label{eqADMM2} \\
  \iter{\tilde{\vect{h}}}{q+1}_{\ing{l},:} & = \vect{v}_{\ing{l},:} \ast \iter{\vect{a}}{q+1} \label{eqADMM3} \\
  \iter{\mtrx{U}}{q+1} & = \iter{\mtrx{U}}{q} + \iter{\tilde{\mtrx{H}}}{q+1} - \iter{\mtrx{H}}{q+1} \label{eqADMM4},
\end{align}
where $\tilde{\vect{h}}_{\ing{l},:}$ and $\vect{u}_{\ing{l},:}$ denote the $\ing{l}$\textsuperscript{th} rows of the matrices $\mtrx{\tilde{H}}$ and $\mtrx{U}$, respectively, while $\Xi$ is the set of all real $(\ing{L}+1)^2 \times \ing{J}$ matrices for which first row has non-negative entries, while the columns $\vect{h}_{:,\ing{j}}$ indexed by $\ing{j} \notin [0,\ing{j}_{\max}]$ contain only zeros. \revis{The mixed norm $\norm{\cdot}{2,1}$ is equal to the sum of the $\ell_2$-norms of matrix columns, while $\norm{\mtrx{H}}{F}$ denotes the Frobenius norm of a matrix $\mtrx{H}$. }

Let $\mathcal{P}_{\Xi}(\mtrx{H})$ denote the operator that projects a matrix $\mtrx{H}$ to $\Xi$, \emph{i.e.}, sets to zero all $h_{0,\forall \ing{j}} < 0$ and $\vect{h}_{:,\ing{j}}$, $\ing{j} \notin [0,\ing{j}_{\max}]$. Define a group soft-thresholding \cite{bach2012optimization} operator $\mathcal{S}_{\mu}(\cdot)$ as
\begin{equation}
  \mathcal{S}_{\mu}(\mtrx{H})_{\ing{l,j}} = \max\left(0, 1 - \frac{\mu}{\norm{\vect{h}_{:,\ing{j}}}{2}} \right)h_{\ing{l,j}},
\end{equation}
where $h_{\ing{l,j}}$ is an entry of the matrix $\mtrx{H}$ at the row $\ing{l}$ and the column $\ing{j}$. Then, the solution of the subproblem \eqref{eqADMM1} is 
\begin{equation}\label{eqProjGroupShrink}
  \iter{\mtrx{H}}{q+1} = \mathcal{S}_{\mu}\left(\mathcal{P}_{\Xi}\left( \iter{\tilde{\mtrx{H}}}{q} + \iter{\mtrx{U}}{q} \right)\right).
\end{equation}
\revis{The time complexity of the above operations is linear, \emph{i.e.} of the order $O(\ing{(L+1)}^2\ing{J})$.}

The subproblem \eqref{eqADMM2} is very similar to the initial constrained quadratic problem \eqref{eqOptim_a}, and can be rewritten as: 
\begin{equation}\label{eqOptim_a_ADMM}
  \min_{\vect{a}} \sum\limits_{\ing{l}=0}^{(\ing{L}+1)^2-1} \sum\limits_{\ing{j}} \left( (v_{\ing{l},:} \ast a)_{\ing{j}} - d_{\ing{l},\ing{j}} \right)^2, \; \; \text{s.t.} \; \; a_0=1,
\end{equation} 
where $d_{\ing{l},\ing{j}} = \iter{h}{q+1}_{\ing{l},\ing{j}} - \iter{u}{q}_{\ing{l},\ing{j}}$. It can again be cast into a linear system, similar to \eqref{eqLPsolution}:
\begin{equation}\label{eqLP_ADMM}
  \sum_{\ing{j}=1}^{\ing{j}_{\max}} a_{\ing{j}} r(\ing{j},\ing{s}) = r_{dv}(0,\ing{s}) -r(0,\ing{s}),
\end{equation}
where $r(\ing{j},\ing{s})$ are defined as in \eqref{eqLPcoeffs}, while the coefficients $r_{dv}(0,\ing{s})$ correspond to the cross-correlation 
\begin{equation}\label{eqLPcoeffs_ADMM}
  r_{dv}(0,\ing{s}) = \sum\limits_{\ing{l}=0}^{(\ing{L}+1)^2-1} \sum\limits_{\ing{j}'} d_{\ing{l},\ing{j}'} v_{\ing{l},\ing{j'-s}}.
\end{equation}

In both cases, the summation with respect to $\ing{j}'$ is now done over the entire range $[-{\ing{J}/2}+1,-{\ing{J}/2}]$. \revis{This means that the autocorrelation coefficients are directly obtained from the power spectrum of GFVV, \emph{i.e.},
\begin{equation}\label{eqACcoeffs_ADMM}
r(\ing{j},\ing{s}) = \mathcal{F}^{-1}\left( \sum\limits_{\ing{l}=0}^{(\ing{L}+1)^2-1} | \hat{\vect{v}}_{\ing{l},:}|^2 \right)_{\ing{j}-\ing{s}}.  
\end{equation}
Since $r(\ing{j},\ing{s})$ is constant across iterations, the Toeplitz matrix and the autocorrelation part of the right hand side of \eqref{eqLP_ADMM} need to be calculated only once.}

\revis{Analogous to autocorrelation, the cross-correlation values \eqref{eqLPcoeffs_ADMM} can be computed in frequency domain from the cross-spectra of the involved quantities at $O((\ing{L}+1)^2 \ing{J}\log{\ing{J}} )$ cost:
\begin{equation}\label{eqCCcoeffs_ADMM}
    r_{dv}(0,\ing{s}) = \mathcal{F}^{-1}\left( \sum\limits_{\ing{l}=0}^{(\ing{L}+1)^2-1} \hat{\vect{d}}_{\ing{l},:} \hat{\vect{v}}^*_{\ing{l},:} \right)_{\ing{s}}.
\end{equation}}

\revis{Even though we have $\ing{j}_{\max} < \ing{J}/2$ (cf. the discussion in subsection \ref{ssec:estimation}), $\ing{j}_{\max}$ and $\ing{J}$ are still comparable, hence the per-iteration complexity of the ADMM algorithm is determined by the cost of solving the linear system \eqref{eqLP_ADMM}, which requires $O((\ing{j}_{\max}+1)^2)$ operations.}

In practice, the convergence speed depends on the parameter $\mu$, which is set to $0.1$ in our experiments. Convergence criterion based on the primal and dual updates has been presented in \cite{boyd2011distributed}, however, we observe that the algorithm typically produces meaningful results within tens of iterations. Furthermore, the algorithm can be accelerated by warm-starting, \emph{i.e.}, by initializing the iterations with the estimates from the previous frame.
}


\bibliographystyle{IEEEbib}
\bibliography{refs}

\end{document}